\documentclass[acmtog]{acmart}

\AtBeginDocument{%
  \providecommand\BibTeX{{%
    \normalfont B\kern-0.5em{\scshape i\kern-0.25em b}\kern-0.8em\TeX}}}

\setcopyright{rightsretained}
\acmJournal{TOG}
\acmYear{2022}\acmVolume{41}\acmNumber{4}\acmArticle{94}\acmMonth{7} \acmDOI{10.1145/3528223.3530110}

\acmConference[Conference acronym 'XX]{Make sure to enter the correct
  conference title from your rights confirmation emai}{June 03--05,
  2018}{Woodstock, NY}
\acmBooktitle{Woodstock '18: ACM Symposium on Neural Gaze Detection,
 June 03--05, 2018, Woodstock, NY} 
\acmPrice{15.00}
\acmISBN{978-1-4503-XXXX-X/18/06}

\usepackage[ruled]{algorithm2e} % For algorithms

\SetAlFnt{\small}
\SetAlCapFnt{\small}
\SetAlCapNameFnt{\small}
\SetAlCapHSkip{0pt}

\usepackage{subfigure}
\usepackage{caption}
\usepackage{algorithmic}
\usepackage{colortbl}

\usepackage{amsmath,amsfonts,bm}

\def\eqref#1{equation~\ref{#1}}
\def\1{\bm{1}}

\def\rva{{\mathbf{a}}}

\def\rvd{{\mathbf{d}}}

\def\rvg{{\mathbf{g}}}
\def\rvh{{\mathbf{h}}}
\def\rvu{{\mathbf{i}}}

\def\rvs{{\mathbf{s}}}

\def\rvu{{\mathbf{u}}}
\def\rvv{{\mathbf{v}}}

\def\rvx{{\mathbf{x}}}

\def\rvz{{\mathbf{z}}}

\def\rmZ{{\mathbf{Z}}}

\DeclareMathAlphabet{\mathsfit}{\encodingdefault}{\sfdefault}{m}{sl}
\SetMathAlphabet{\mathsfit}{bold}{\encodingdefault}{\sfdefault}{bx}{n}

\newcommand{\expec}{\mathbb{E}}

\acmJournal{TOG}

\begin{document}

\title{ASE: Large-Scale Reusable Adversarial Skill Embeddings for Physically Simulated Characters}
%%SL.1.1: Some title ideas (I'm not *thrilled* with any of these, but maybe something here will give you some ideas):
% [METHODNAME]: Self-Supervised Reinforcement Learning of Reusable Skill Spaces
% [METHODNAME]: Flexible Animation with Self-Supervised Skill Spaces
% [METHODNAME]: Reinforcement Learning of General Controllers from Large Datasets
% Some method name ideas:
% ARC - Adversarial Reusable Controllers
% RAMP - Resusable Adversarial Motion Priors :)

\author{Xue Bin Peng}
\email{xbpeng@berkeley.edu}
\orcid{0002-3677-5655}
\affiliation{
  \institution{University of California, Berkeley}
  \country{USA}
}
\affiliation{
  \institution{NVIDIA}
  \country{Canada}
}

\author{Yunrong Guo}
\email{kellyg@nvidia.com}
\affiliation{
  \institution{NVIDIA}
  \country{Canada}
}

\author{Lina Halper}
\email{lhalper@nvidia.com}
\affiliation{
  \institution{NVIDIA}
  \country{Canada}
}

\author{Sergey Levine}
\email{svlevine@eecs.berkeley.edu}
\affiliation{
  \institution{University of California, Berkeley}
  \country{USA}
}

\author{Sanja Fidler}
\email{lhalper@nvidia.com}
\affiliation{
  \institution{University of Toronto}
  \country{Canada}
}
\affiliation{
  \institution{NVIDIA}
  \country{Canada}
}


%%SL.1.24: Is there some formatting error? I don't see the word "Abstract" in the PDF, but maybe the TOG format changed so that this word doesn't appear anymore? Maybe just double check that the style is correct
%%JP: this should be correct, there's no "abstract" label in the style.
\begin{abstract}
The incredible feats of athleticism demonstrated by humans are made possible in part by a vast repertoire of general-purpose motor skills, acquired through years of practice and experience. These skills not only enable humans to perform complex tasks, but also provide powerful priors for guiding their behaviors when learning new tasks. This is in stark contrast to what is common practice in physics-based character animation, where control policies are most typically trained from scratch for each task. In this work, we present a large-scale data-driven framework for learning
%%SL.1.24: maybe use a different word than "developing"? developing kind of sounds like you're coding it up...
%%JP: switched to learning
versatile and reusable skill embeddings for physically simulated characters. Our approach combines techniques from adversarial imitation learning and unsupervised reinforcement learning to develop skill embeddings that produce life-like behaviors, while also providing an easy to control representation for use on new downstream tasks. Our models can be trained using large datasets of unstructured motion clips, without requiring any task-specific annotation or segmentation of the motion data. By leveraging a massively parallel GPU-based simulator, we are able to train skill embeddings using over a decade of simulated experiences, enabling our model to learn a rich and versatile repertoire of skills. We show that a single pre-trained model can be effectively applied to perform a diverse set of new tasks. Our system also allows users to specify tasks through simple reward functions, and the skill embedding then enables the character to automatically synthesize complex and naturalistic strategies in order to achieve the task objectives.
\end{abstract}




\begin{teaserfigure}
  \includegraphics[width=1\textwidth]{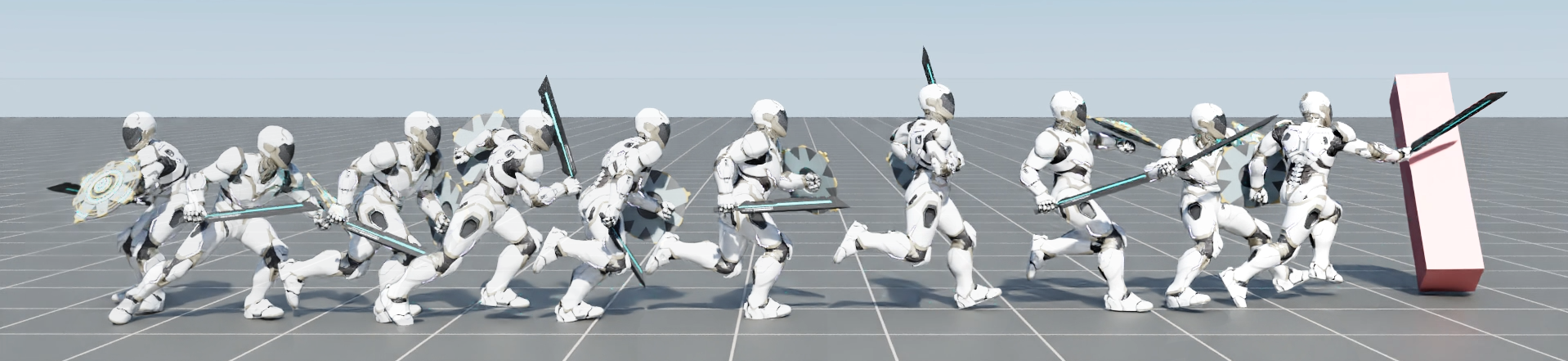}
  \caption{Our framework enables physically simulated characters to learn versatile and reusable skill embeddings from large unstructured motion datasets, which can then be applied to produce life-like behaviors for new tasks. Here, a character is utlizing behaviors from a learned skill embedding in order to run to a target and knock it over.}
  %%SL.1.24: I think we can have a more inspiring caption. It would be great if the caption could provide a few sentences to basically summarize the main idea. Don't worry if the caption is a little long -- often this is the very first thing reviewers will see and read (even before the abstract), so it's worth spending a little bit of space on it to make a great first impression.
  %%JP: ok, added more info to the caption, hopefully this is a bit better
  \label{fig:teaser}
\end{teaserfigure}

\maketitle

\section{Introduction}
Humans are capable of performing an awe-inspiring variety of complex tasks by drawing on a vast repertoire of motor skills. This repertoire is built up over a lifetime of interaction with the environment, leading to general-purpose skills that can be widely reused to accomplish new tasks.
%%SL.1.1: Generally, the above sentences are reasonable, and I see what you are driving toward -- that people have lots of different skills to draw on from their past experience. But I think the above two sentence should be refined a bit. It's not clear what "awe-inspiring feats of agility" has to do with "vast repertoire" -- indeed, it seems like there are two different axes, one of "agility" and one of "generality," and it's not obvious that the two are actually related (rather than orthogonal), so some people might not buy what you are saying up above.
%JP: ok, tweaked
This is at odds with what is conventional practice in physics-based character animation and reinforcement learning, where control policies are typically trained from scratch, to specialize in a single specific task. Developing more versatile and reusable models of motor skills can enable agents to solve tasks that would otherwise be prohibitively challenging to learn from scratch. However, manually constructing a sufficiently rich set of tasks and reward functions that can give rise to behaviors that are as diverse and versatile as those of humans would require an immense engineering effort.

How then can we endow agents with large and versatile repertoires of skills? Some inspiration may be drawn from domains such as computer vision and natural language processing, where large expressive models trained on massive datasets have been a central component of major advances \cite{HinSal06,deng2009imagenet,word2vec2013,VAEKingmaW13,Doersch2015,resnet2016,oord2018cpc,GPT32020,SelfSupvervisedJing2021}. Not only can these models solve challenging tasks, but they also provide powerful priors that can be reused for a wide range of downstream applications \cite{pmlr-v32-donahue14,vinyals2014neural,Sermanet2017TCN,GPT32020,TransferRaffel2020,promptuning2021}. A similar data-driven paradigm may also be applicable for developing more general and versatile motor skills. Rather than laboriously designing a rich set of training tasks that leads to a flexible range of behaviors, we can instead provide the agent with a large unstructured motion dataset, containing examples of behaviors that we would like the agent to acquire. The agent can then be trained to perform a large variety of skills by imitating the behaviors depicted in the dataset.
%%SL.1.9: The above paragraph reads well, but the sentence below is a bit confusing. Perhaps some of this is inevitable because we can't describe our entire approach in one sentence, but perhaps there is some way to be more explicit here, e.g.: If this process can be set up so as to extract a suitable representation of feasible and realistic behaviors, downstream tasks can then be learned by selecting among these available behaviors, which would constrain them to naturalistic and useful motions [or something like that]
By modeling the learned skills with representations that are suitable for reuse, we may then be able to develop more capable and versatile motor skill models that can be re-purposed for a wide range of new applications.

%%SL.1.9: Maybe there is a cleaner transition we can have, like "To this end, we present..." [basically to make it clear that what we present aims to accomplish the thing we motivated in the previous para]
%%JP: added
To this end, we present adversarial skill embeddings (ASE), a scalable data-driven approach for learning general and reusable motor skills for physically simulated characters. Given a large dataset of unstructured motion clips, our system trains a low-level latent variable model to produce behaviors that resemble those shown in the dataset. Importantly, the learned skills need not exactly match any particular motion clip. Instead, our pre-training objective encourages the model to learn a diverse set of skills that exhibits the general behavioral characteristics depicted in the dataset. This is enabled by an information maximization objective, which spurs the model to discover diverse and distinct behaviors. Once trained, the low-level model can then be used to define an abstract action space for a high-level policy to perform new downstream tasks. By pre-training the low-level model with motion clips of naturalistic behaviors recorded from human actors, the model can be used to synthesize agile and life-like behaviors (see Figure~\ref{fig:teaser}), without requiring any additional motion data or extensive reward engineering.
%%SL.1.9: It would be great to reference a diagram or figure of some sort in the above paragraph.
%%JP: ok, added a ref to the teaser, and will add some images of the character performing some tasks there once we have some nice renderings

The central contribution of this work is a scalable adversarial imitation learning framework that enables physically simulated characters to acquire large repertoires of complex and versatile motor skills,
%%SL.1.1: As per above, the phrasing "autonomous acquire" is somewhat misleading. Perhaps better to emhpasize universality, generality, and reusability, rather than autonomous acquisition or unsupervised
%%JP: removed "autonomous" and changed "self-supvervised RL" to "adversarial imitation learning"
which can then be reused to perform a wide range of downstream tasks. Our system is able to learn behaviors from large unstructured motion datasets, containing over a hundred diverse motion clips. By leveraging Isaac Gym, NVIDIA's massively parallel GPU simulator,
%%SL.1.1: Not sure this belongs here, maybe more of an implementation detail that goes in the experiments section? (the test is, was this required for the method to work? my guess is no)
%%JP: I think being able to scale up training a lot is vital to getting this to work in practice. I don't think we would have been able to train these models in a reasonable amount of time using just a regular CPU simulator.
%%SL.1.9: I agree that scale is a big part of it, I just meant that saying "we did it using Isaac Gym" sounds kind of weird and mysterious (also kind of comes off as an advertisement) -- basically, if some aspect of scale is important, we should lead by saying what it was that was actually important (e.g., parallelized simulation or something), and only then mention which particular package was used, as otherwise it's kind of putting the cart before the horse.
%%JP: ok, replace mention of isaac gym with "massive parallel GPU simulator", and will mention isaac gym more explicitly later in the experiments section
%%JP: sanja suggests that we mention nvidia explicitly, so adding that back
%%SL.1.24: hmm ok, I would be a little nervous it might rub reviewers the wrong way (I don't think it rises to the level of anonymity violation, just some people might get annoyed if they view it as less relevant and more marketing-focused), but I think it's up to you guys how to handle it, it's probably not a big deal in the end
%%JP: lets leave this in for now, its probably no worse than mention bullet or mujoco explicitly
our system is able to pre-train general-purpose motor skill models using the equivalent of a decade of simulated experiences. We propose a number of important design decisions that increases the diversity of the skills acquired during the pre-training process, and improve the overall effectiveness of the models when transferred to downstream tasks. Furthermore, by pre-training the model to recover from random initial states, we can develop highly robust recovery strategies that can agilely and consistently recover from substantial external perturbations. Once trained, these recovery strategies can then be seamless integrated into new tasks, without any additional algorithmic overhead. Code and data for this work is available at \href{https://xbpeng.github.io/projects/ASE/index.html}{https://xbpeng.github.io/projects/ASE/}.
%%SL.1.1: A few general things about the intro: (i) I think we need to be more explicit in stating the problem statement -- take lots of data, learn a latent space of behaviors, and then solve downstream tasks by learning policies that navigate this latent space, where the space contains all the "good" things the character could do; (ii) We need a teaser figure + diagram; (iii) Focusing more on scalable learning and using lots of data to get a universal controller is probably more compelling (and less likely to be perceived as misleading) than focusing on unsupervised RL.
%%JP: ok, rewrote the intro to focus on learning more general motor skills models from large datasets
%%JP: will add a teaser a bit later

\section{Related Work}
Developing controllers that can produce agile and life-like behaviors has been one of the core challenges of computer animation \cite{Raibert1991,AthleticsHodgins1995,Zordan2002,Yin07,2010-TOG-gbwc}. Optimization techniques, based on trajectory optimization or reinforcement learning, are some of the most widely used methods for synthesizing controllers for simulated characters \citep{Panne94virtualwindup,Yin08,delasa2010,CIO2012,pmlr-v28-levine13,BicycleTan2014}. These techniques synthesize controllers by optimizing an objective function, which encodes properties of a desired skill. While these methods are able to synthesize physically plausible motions for a wide variety of behaviors \cite{BipedEndo2005,TanSwimming2011,Tassa2012,AlBorno2013,Wampler2014,Gehring2016}, designing effective objectives that lead to naturalistic motions can often involve a tedious and labour-intensive development process. Heuristics derived from prior knowledge regarding the characteristics of natural behaviors can be incorporated into the objective in order to improve motion quality, such as enforcing energy efficiency, lateral symmetry, stability, and many more \citep{Geyer2173,BipedWang2009,CIO2012,SymYu2018}. However, these heuristics are generally not broadly applicable for all skills, and different skills often require different carefully curated sets of heuristics in order to produce life-like behaviors. Incorporating more biologically accurate simulation models can also improve motion quality \citep{MuscleWang2012,2013-TOG-MuscleBasedBipeds,MuscleJiang2019}, but may nonetheless produce unnatural behaviors without the appropriate objective functions \cite{Song2020}.

\paragraph{Data-driven methods:}
The difficulty of designing controllers and objective functions that produce naturalistic behaviors has motivated the widespread adoption of data-driven methods for physics-based character animation \citep{Zordan2002,Sharon2005Walking,DaSilva2008,DataDrivenLee2010,PendulumKwon2017,GANPei2021}. These techniques can produce highly life-like motions by imitating reference motion data. Many of these methods utilize some form of motion tracking, where controllers imitate reference motions by explicitly tracking the sequence of target poses specified by a motion clip \citep{BipedDataSok2007,DataDrivenLee2010,Liu2010Samcon,2016-TOG-controlGraphs,2018-TOG-deepMimic,Baskeball2018}.
%Additional task objectives can be incorporated into the training process to enable characters to perform auxiliary tasks that may not be depicted in the original motion data \cite{2018-TOG-deepMimic,Lee2021Parameterized}. However, the tracking objective tend to prevent the character from deviating significantly from the reference motion, which in turn can limit the flexibility of the character to satisfy additional task objectives.
However, it can be difficult to apply tracking-based techniques to imitate behaviors from large and diverse motion datasets. Composition of disparate skills often requires some form of a motion planner to select the appropriate motion for the character track in a given scenario \cite{2012-TOG-TerrainRunner,2017-TOG-deepLoco,PredictSimPark2019,DreCon2019}, which can be difficult to construct for complex tasks. More recently, \citet{2021-TOG-AMP}
%%SL.1.1: Are there any other GAN-based imitation methods we should cite in the context of this?
proposed adversarial motion priors, which allow characters to perform tasks while imitating behaviors from large unstructured motion datasets. This enables characters to automatically compose and sequence different skills, without being constrained to explicitly track a particular motion clip.
%%SL.1.9: It kind of feels like the above sentence doesn't complete the idea -- it kind of talks about how AMP is great, but then in the next sentence you change topics without saying what the greatness of AMP has to do with the method in this paper (e.g., is ASE better somehow? or does it benefit from this somehow?), leaving the reader wondering what the above discussion has to do with our method here.
While these motion imitation methods can produce high quality results, they predominantly train models from scratch for each task.
%%SL.1.1: "predominantly" implies that there are some that don't -- do we discuss those anywhere? that seems pretty relevant...
%%JP: the hierarchical methods which reuse pre-trained skills are covered in the "hierarchical models" section
This \emph{tabula-rasa} approach can incur significant drawbacks in terms of sample efficiency and can limit the complexity of tasks that characters can accomplish, requiring agents to relearn common behaviors, such as walking, over and over again for each new task. Our work aims to learn reusable motor skills from large motion dataset, which can then be leveraged to synthesize naturalistic behaviors for new tasks without requiring additional motion data or retraining from scratch each time.
%%SL.1.9: Maybe just make explicit how this relates to our method by adding one sentence stating how ours is different?
%%JP: added a sentence at the end for this

\paragraph{Hierarchical models:}
One way to reuse previously acquired skills for new tasks is by building a hierarchical model. This approach generally consists of two stages: a pre-training stage, where a collection of controllers are trained to specialize in different low-level skills (e.g., walking in different directions) \cite{Coros09,2012-TOG-TerrainRunner,HeessWTLRS16,hausman2018learning,TwoPlayerWon2021}, and a task-training stage, where the low-level controllers are integrated into a control hierarchy with a high-level controller that leverages the low-level skills to perform new downstream tasks (e.g., navigating to a target location) \cite{FaloutsosComposable2001,YeAbstractModel2010,MordatchPlanning2010,2017-TOG-deepLoco,ControlFrags2017,2020-TOG-MVAE}. General motion tracking models trained to imitate reference motion data can be used as effective low-level controllers \cite{MocapImitationChentanez2018,wang2020unicon,ScalableWon2020}. To apply these tracking models to downstream tasks, a high-level kinematic motion planner can be used to specify target motion trajectories for guiding the low-level controller towards completing a desired task \cite{DreCon2019,PredictSimPark2019}. However, constructing such a motion planner typically requires maintaining a motion dataset for use on downstream tasks. Latent variable models trained via motion tracking can obviate the need for an explicit motion planner by allowing the high-level controller to steer low-level behaviors via latent variables \cite{RobustImitation2017,pmlr-v100-lynch20a,CatchCarry2020,merel2018neural,MCPPeng19,CARLLuo2020,ComicHasenclever2020}.
%%SL.1.1: Since our approach is also a latent variable model, do we need to elaborate on how it relates to these?
%%JP: most of these latent variable models are trained using motion tracking. The limitations of that, and how our method is different is discussed below
%%SL.1.9: OK, I think that's a good argument in favor of our approach, but I think this part should be fleshed out a little. First, it is left somewhat implicit what the difference between these prior latent var methods and our method is (you say adversarial imitation, but readers might not know what that means as far as the practical difference). Second, the current phrasing kind of makes it sound like this is true for *all* prior latent var methods. If it's only true for most, then probably we need discussion of the exceptions? We discussed the prior work stuff in detail before, but it doesn't seem like this discussion ended up in the writing here... it would be good to be very clear on what the closest works are and precisely how they are different, as these latent var things are the most likely source of trouble during reviewing.
While motion tracking can be an effective method for constructing low-level controllers, the tracking objective can ultimately limit a model's ability to produce behaviors that are not depicted in the original dataset, thereby potentially limiting a model's flexibility to develop more general and versatile skills. In this work, instead of using motion tracking, we present a more flexible pre-training approach that combines adversarial imitation learning and unsupervised reinforcement learning,
%%SL.1.1: see my prior comment about the term "unsupervised"
%%JP: changed to "adversarial imitation learning"
which provides agents more flexibility in developing novel and versatile behaviors.

%%SL.1.1: My sense is that if we present our method not (primarily) as an unsupervised approach but rather as a way to utilize lots of data, then this paragraph can explain in more detail how it is similar to and different from *actual* unsupervised RL methods
\paragraph{Unsupervised reinforcement learning:}
In the unsupervised reinforcement learning setting, agents are not provided with an explicit task objective. Instead, the goal is for the agent to learn skills autonomously by optimizing an \emph{intrinsic} objective derived from the agent's own past experiences. These intrinsic objectives typically motivate the agent to seek novelty or diversity, which can be quantified using surrogate metrics, such as errors from model predictions  \cite{StadieLA15,AchiamS17,pathak17a,burda2018exploration,pathak19a}, state visitation counts \cite{CountExpStrehl2008,ExpBellemare2016,Ex2Fu2017,ExpTang2017,FlorensaDA17}, or entropy of the agent's state distribution \cite{hazan19a,APTLiu2021}. In this work, we will leverage a class of techniques based on maximizing mutual information between abstract skills and their resulting behaviors \cite{VICGregor2017,eysenbach2019diversity,APSLiu2021,SharmaGLKH20dads,RVICBaumli2020}. While unsupervised reinforcement learning has shown promising results on relatively low-dimensional problems, when applied to more complex settings with large numbers of degrees-of-freedom, these unsupervised reinforcement learning techniques often fail to discover useful behaviors. Furthermore, without additional prior knowledge, it is unlikely that unsupervised reinforcement learning techniques alone will discover naturalistic, life-like behaviors that resemble the motions of humans or animals. Therefore, our models are trained using a combination of an adversarial imitation learning objective and an unsupervised information maximization objective.
%which encourages a model to develop diverse and distinct skills, while also adopting behaviors that resemble those shown in the motion dataset.
Adversarial imitation learning allows our models to mimic realistic motions from user-provided data, while unsupervised reinforcement learning techniques allow the model to learn skill representations that are more directable and mitigate the common problem of mode-collapse associated with adversarial training methods by promoting more diverse behaviors.
%%SL.1.9: Generally I think the above paragraph is written quite well, but it does seem to *slightly* miss the point, in that the role of data is not only to make unsupervised RL work better, but also to actually make the motions look realistic -- that is, unsupervised RL wouldn't result in realistic motions no matter how well it works.
%%JP: added a sentence to say that URL will not be able to discover life-like motions on its own

%%SL.1.1: I wonder if this section would be more exciting if we call it "Problem Formulation" and present it kind of as a counterpoint to the classic way of learning skills with RL? Basically, instead of making this seem like a system overview, use this as a way to present the basic idea that we can learn latent space from lots of data and then use it to solve many tasks, and motivate very clearly *why* this is a good idea (the current paragraph is already doing about 90% of this, so it seems like we're nearly there)

\section{Reinforcement Learning Background}
%%SL.1.1: Right now the background section is very long, which defers the interesting novel stuff until a lot later. That seems non-ideal. I wonder if we can reduce the length of this as much as possible, and maybe not even have a "unsupervised RL" subsection at all, just a quick primer on the basics?
%%JP: removed the URL part and just kept a brief overview of RL
In our framework, both pre-training and transfer tasks will be modeled as reinforcement learning problems, where an agent interacts with an environment according to a policy $\pi$ in order to optimize a given objective \cite{Sutton1998}. At each time step $t$, the agent observes the state $\rvs_t$ of the system, then samples an action from a policy $\rva_t \sim \pi(\rva_t | \rvs_t)$. The agent then executes the action, which leads to a new state $\rvs_{t+1}$, sampled according to the dynamics of the environment $\rvs_{t+1} \sim p(\rvs_{t+1} | \rvs_t, \rva_t)$, and a scalar reward $r_t = r(\rvs_t, \rva_t, \rvs_{t+1})$. The agent's objective is to learn a policy that maximizes its expected discounted return $J(\pi)$,
\begin{equation}
    J(\pi) = \expec_{p(\tau | \pi)} \left[ \sum_{t=0}^{T-1} \gamma^t r_t \right],
\label{eqn:rl_objective}
\end{equation}
where $p(\tau | \pi) = p(\rvs_0) \prod_{t=0}^{T-1} p(\rvs_{t+1} | \rvs_t, \rva_t) \pi(\rva_t | \rvs_t)$ represents the likelihood of a trajectory $\tau = \{\rvs_0, \rva_0, r_0, \rvs_1, ..., \rvs_{T-1}, \rva_{T-1}, r_{T-1}, \rvs_T \}$ under $\pi$. $p(\rvs_0)$ is the initial state distribution, $T$ denotes the time horizon of a trajectory, and $\gamma \in [0, 1]$ is a discount factor. The reward function thus provides an interface through which users can specify the task that an agent should perform. However, designing effective reward functions that elicit the desired behaviors from an agent often involves a tedious design process. Therefore, constructing a sufficiently rich set of tasks that leads to diverse and sophisticated motor skills can present a daunting engineering endeavour.

\section{Overview}
\begin{figure}[t]
	\centering
    \includegraphics[width=1\linewidth]{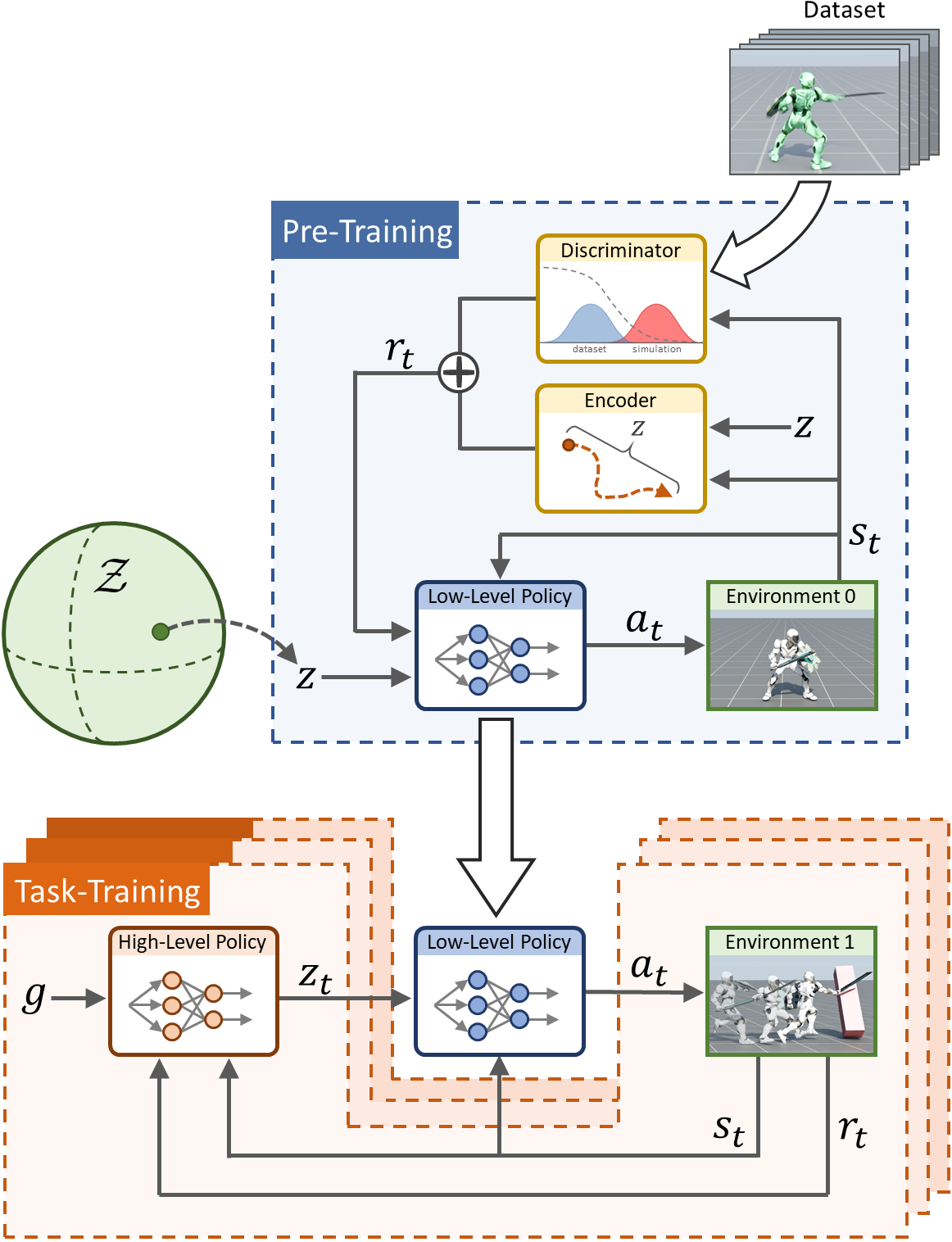}
    %%SL.1.24: Generally, I think this figure looks very nice, but something it doesn't quite do very clearly is illustrate that there is a *space* of behaviors that is learned. I.e., there is never any icon or picture representing the "z space" or indicating that it is in some sense the link between the high level and low level. Maybe that's OK, as the picture is quite detailed already....
    %%JP: added a visualization for the latent space, the figure might be becoming a bit too busy tho
    \caption{The ASE framework consists of two stages: pre-training and transfer. During pre-training a low-level policy $\pi(\rva | \rvs, \rvz)$ is trained to map latent skills $\rvz$ to behaviors that resemble motions depicted in a dataset. The policy is trained to model a diverse repertoire of skills by using a reward function that combines an adversarial imitation objective, specified by a discriminator $D$, and an unsupervised skill discovery, specified by an encoder $q$. After pre-training, $\pi$ can be transferred to new tasks by using a task-specific high-level policy $\omega(\rvz | \rvs, \rvg)$ to specify latent variables $\rvz$ for directing the low-level policy towards accomplishing a task-specific goal $\rvg$.}
\label{fig:overview}
\end{figure}

In this paper, we present adversarial skill embeddings (ASE), a scalable data-driven approach for learning reusable motor skills for physically simulated characters.
%%SL.1.1: Somewhere here would be a good place to reference a figure...
%%JP: added an overview figure
An overview of our framework is provided in Figure~\ref{fig:overview}.
%%SL.1.9: Having a figure like this is great, but I think the current figure is a little bit too abstract (though broadly speaking it has the right idea) -- is there a way to make it more visual somehow with some icons that explain the parts in a more self-explanatory fashion? E.g., pictures of characters doing various tasks in the top box and some diagrams of characters doing more task-directed things in the bottom box? Maybe also make it a bit more obvious what the various symbols like omega mean
%%JP: ok, added some visualization for each component. Will add an image for the environment once the renderings are available.
The training processes consists of two stages: a pre-training stage, and a task-training stage. During pre-training, the character is given a dataset of motion clips $\mathcal{M} = \{m^i\}$, which provides examples of the kinds of behaviors that the agent should learn.
%%SL.1.9: This generally seems reasonable, but we should be very precise with what exactly we mean here -- is it "behavioral style" or just examples of naturalistic motion or something else? Basically, there is some assumption being made about what is actually in the dataset, and it would be good to somehow be explicit about this dataset is really supposed to be specifying (and also what it is *not* supposed to be specifying).
%%JP: reworded it as "the kinds of behaviors that the agent should learn". Hope that is more clear.
%%SL.1.24: I think that's better, but maybe we should add a little bit more discussion about what we are assuming (or not assuming) about this dataset, e.g. something like: This dataset could be very large, and might simply consist of all motion data that is consistent with the style that the user would like to see, regardless of whether it is directly relevant for particular downstream tasks or not. In our experiments, we use [whatever]. [this is I think important to point out, because some readers will have the wrong idea about what assumptions we do or do not make, since this is something that distinguishes our work from prior work in some sense]
Each motion clip $m^i = \{\rvs_t^i\}$ is represented as a sequence of states that depicts a particular behavior. This dataset is used by an adversarial imitation learning procedure to train a low-level skill-conditioned policy $\pi(\rva | \rvs, \rvz)$, which maps latent variables $\rvz$ to behaviors that resemble motions shown in the dataset. Unlike most prior motion imitation techniques \cite{DataDrivenLee2010,Liu2010Samcon,2018-TOG-deepMimic,CatchCarry2020},
%%SL.1.9: citations
%%JP: added
our models need not exactly match any particular motion clip in the dataset. Instead, the objective is for a policy to discover a diverse and versatile set of skills that exhibits the general characteristics of the motion data. Then in the task-training stage, the low-level policy is reused to perform new tasks by training a task-specific high-level policy
%%SL.1.9: maybe clearer and more conventional to denote the high-level as \pi_something rather than \omega? could even be \pi_\omega or \pi_\text{hi} or something
$\omega(\rvz | \rvs, \rvg)$, which specifies latent variables for directing the low-level policy towards completing the task objectives. The high-level policy is conditioned on a task-specific goal $\rvg$, and can be trained without any motion data. But since the low-level policy was trained with a motion dataset, it allows the character to produce naturalistic behaviors even in the absence of motion data.

%%SL.1.1: Right here, it might be a little difficult for readers to figure out where to place this in the context of your method. The "Overview" section didn't really provide a system overview but rather a problem statement. That's OK, but that will leave readers here a bit unsure about what the pieces are and how they fit together. Given that this is the first real technical section of the paper, I think it would help to start it with some discussion of technical background to basically explain to the reader how the parts fit and where the part you are describing now will fit in. A figure/diagram would help with this. Additionally, a section title that more clearly signals that now you'll be describing your thing would also be really helpful (maybe a top-level section with the same name as the method?)
\section{Adversarial Skill Embeddings}
\label{sec:ASE}

%%SL.1.9: I wonder if the flow will be smoother if we do background, then overview, then this (in fact overview could just be the first paragraph of this section) -- as is the beginnign of this kind of re-motivates some of the stuff that already happened in the overview
%%JP: moved
In ASE, the skills are modeled by a skill-conditioned policy $\pi(\rva | \rvs, \rvz)$, where each skill is represented by a latent variable $\rvz \in \mathcal{Z}$ sampled according to a prior over latent skills $\rvz \sim p(\rvz)$. An effective skill model should produce realistic behaviors, while also providing a \emph{directable} skill representation that can be conveniently utilized to perform new downstream tasks. This is accomplished by combining a motion imitation objective with an unsupervised skill discovery objective, which encourages the agent to develop a set of diverse skills that conforms to the behavioral characteristics specified by a dataset of motion clips. Given a motion dataset $\mathcal{M}$, the pre-training objective is given by:
\begin{equation}
    \mathop{\mathrm{max}}_\pi \quad  - D_\mathrm{JS}\left(d^\pi(\rvs, \rvs') \middle| \middle| d^\mathcal{M}(\rvs, \rvs') \right) + \beta \ I\left(\rvs, \rvs' ; \rvz \middle| \pi \right) .
    \label{eqn:sd_objective}
\end{equation}
%%SL.1.1: OK, so the above objective basically represents the full objective of your method for learning the low-level policy. I think it should be presented as such. We should put the MI and the GAN on a more equal footing: instead of presenting this as MI skill discovery with a GAN, I think it's better to motivate kind of like this: if we want a general-purpose motion prior, we need it to produce motions that look realistic, and we need it to be directable, in the sense that a higher level policy can tell it what to do; that's what the objective does: the first part says it should be controllable, and the second part says it should cover the data. If we motivate it in this more balanced way, I think it will be easier for the audience to understand why it's a good idea, otherwise it kind of feels incongruent to be describing an animation system where the data-dependent term (the only thing that makes the motions look good) is kind of an afterthought; the point of the Djs term is not just to overcome the problems of prior skill discovery methods, because our goals are fundamentally just different: we are not trying to produce all the skills, but only the skills that look similar to the data
The first term is the imitation objective, which encourages the policy to produce realistic behaviors by matching the marginal state-transition distribution of the dataset,
%%SL.1.24: Maybe it would be good to stop the sentence here and add one sentence of intuition (since this is so important): Intuitively, this means that a successful low-level policy should be capable of reproducing the \emph{entire} distribution of behaviors illustrated by the training data. Note that this does not mean that the policy simply memorizes the training motions -- it requires executing these motions inside of a physical simulator, and generalizing to the entire distribution from which these motions were sampled, which might entail interpolating the provided behaviors. [or something along these lines]
%%JP: sounds good, added a sentence below to emphasize this intuition
where $D_\mathrm{JS}$ denotes the Jensen-Shannon divergence, $d^\pi(\rvs, \rvs')$ represents the marginal distribution of state transitions induced by $\pi$, and $d^\mathcal{M}(\rvs, \rvs')$ is the marginal state-transition distribution of the motion dataset $\mathcal{M}$. Intuitively, this objective encourages a low-level policy to reproduce the \emph{entire} distribution of behaviors from which the motion data were sampled, rather than simply imitating the individual motion clips. An effective skill embedding, should therefore be able to generalize to behaviors that are not depicted in the motion clips, but still conform to the general characteristics of the dataset. The second term is the unsupervised skill discovery objective, which encourages the policy to develop a diverse and distinct set of skills
%%SL.1.9: maybe good to remind the reader in this sentence (or nearby) *why* we need diverse skills and just matching the distribution is not enough
by maximizing the mutual information between a skill $\rvz$ and the resulting behaviors, as represented by state transitions. $\beta$ is a manually specified coefficient.
%%SL.1.24: Again, it might be good to add a few sentences of intuition, like: Intuitively, this objective encourages the low-level policy to be controllable, such that different values of z lead to different behaviors, and therefore z can be used to ``steer'' the policy into performing different behaviors from the training distribution
%%JP: added some intuition below
This objective encourages the low-level policy to be \emph{directable}, such that different values of $\rvz$ leads to distinct and predictable behaviors.
Computing these quantities exactly is generally intractable. In the following subsections, we will discuss techniques for constructing a practical approximation of this objective.
%%SL.1.9: I wonder, could parts of the above paragraph somehow be related to the overview diagram? In general, it seems like the above paragraph is very important because it essentially explains what we are doing (and everything else is kind of "implementation details"), but right now it goes by pretty fast, and some readers might not get precisely what the parts of our method actually are

%%SL.1.24: One thought here. I don't know if this is necessarily optimal, but a potential issue I might foresee is that some readers might not be convinced from reading the related work section that our method is really different from prior latent space methods. Could it be that after we explain the technical approach for the latent space, we can squeeze in a sentence or two that basically says how all this is different from prior latent space techniques? Of course, if it just makes them look similar, that could do more harm than good, so I'm not sure...

\subsection{Imitation Objective}
\label{sec:imitation_objective}
Since computing the Jensen-Shannon divergence can be computationally challenging,
%%SL.1.24: I feel like we shouldn't put quite so much emphasis on the intractable thing -- it kind of leads the reader into lateral thinking (e.g., why not use another more tractable objective? ofc that's irrelevant, but they'll wonder)
%%JP: is it any better to just say it's challenging?
we will leverage a variational approximation in the form of a GAN-like framework \cite{GAN2014}. First, we introduce a discriminator $D(\rvs, \rvs')$, which is trained to classify if a given state transition $(\rvs, \rvs')$ is from the dataset or was produced by the agent,
\begin{equation}
    \mathop{\mathrm{min}}_D \quad -\expec_{d^\mathcal{M}(\rvs, \rvs')} \left[ \mathrm{log}\left( D\left(\rvs, \rvs' \right) \right) \right] - \expec_{d^\pi(\rvs, \rvs')} \left[ \mathrm{log}\left(1 - D\left(\rvs, \rvs' \right) \right) \right] .
\end{equation}
The policy can then be trained using rewards specified by $r_t = -\mathrm{log}\left(1 - D\left(\rvs_t, \rvs_{t+1} \right) \right)$. It can be shown that this adversarial training procedure minimizes the Jensen-Shannon divergence between $d^\pi(\rvs, \rvs')$ and $d^\mathcal{M}(\rvs, \rvs')$ \cite{fgan2016,ke2021imitation}. This objective is similar to the one presented by \citet{2021-TOG-AMP}, and can be interpreted as training a policy to produce behaviors that appear to the discriminator as being indistinguishable from motions shown in the dataset. But unlike \citet{2021-TOG-AMP}, which leverages a discriminator to shape the behavioral style of policies trained from scratch for a particular task, our framework utilizes a discriminator to learn reusable skill embeddings, which can later be used to perform new tasks. However, simply matching the behavioral distribution of the data, does not ensure that the low-level policy learns a \emph{directable} skill representation that can be effectively reused for new tasks. In the next subsection, we will present a skill discovery objective for acquiring more directable skill representations.

\subsection{Skill Discovery Objective}
\label{sec:skillDiscoveryObjective}
The imitation objective encourages the policy to produce behaviors that resemble the dataset, but it does not ensure that $\pi$ learns a skill representation that is easy to control and amenable for reuse on downstream tasks. Furthermore, the adversarial training procedure described in Section~\ref{sec:imitation_objective} is prone to mode-collapse, where the policy reproduces only a narrow range of behaviors depicted in the original dataset.
%%SL.1.9: On the surface, these seem like two different points made at two different levels of abstraction: the first is that we need skills (very important) the second that GANs have mode collapse. Somehow the second point seems relatively minor and actually detracts from the (very important) first point, unless they can be explicitly related to each other somehow
To address these shortcomings, we incorporate a skill discovery objective, commonly used in unsupervised reinforcement learning, to encourage the policy to model more diverse and distinct behaviors. The objective aims to maximize the mutual information $I(\rvs, \rvs' ; \rvz | \pi)$ between the state transitions $(\rvs, \rvs')$ produced by a policy $\pi$, and a latent skill variable $\rvz$, drawn from a distribution of skills $\rvz \sim p(\rvz)$. The intuition for this choice of objective can be made more apparent if we consider the definition of mutual information,
\begin{equation}
    I(\rvs, \rvs' ; \rvz | \pi) = \mathcal{H}(\rvs, \rvs' | \pi) - \mathcal{H}(\rvs, \rvs' | \rvz, \pi) .
    \label{eqn:mutual_info}
\end{equation}
Maximizing the mutual information can be interpreted as maximizing the \emph{marginal} state-transition entropy $\mathcal{H}(\rvs, \rvs' | \pi)$ induced by $\pi$, while minimizing the \emph{conditional} state-transition entropy $\mathcal{H}(\rvs, \rvs' | \rvz, \pi)$ produced by a particular skill $\rvz$. In other words, the policy should learn a set of skills that produces diverse behaviors, while each individual skill should produce a distinct behavior.

Unfortunately, computing Equation~\ref{eqn:mutual_info} is intractable in most domains. Therefore, we will approximate $I(\rvs, \rvs' ; \rvz | \pi)$ using a variational lower bound presented by \citet{VICGregor2017} and \citet{eysenbach2019diversity}. But unlike these prior methods, instead of learning a discrete set of skills, our policy will be trained to model a continuous latent space of skills, which is more amenable to interpolation and blending between different behaviors. To derive this objective, we first note that determining the marginal entropy $\mathcal{H}(\rvs, \rvs')$ is also generally intractable. A more practical objective can be obtained by taking advantage of the symmetry of mutual information,
\begin{align}
I(\rvs, \rvs' ; \rvz | \pi) & = I(\rvz ; \rvs, \rvs' | \pi) \\
& = \mathcal{H}(\rvz) - \mathcal{H(\rvz | \rvs, \rvs', \pi)} .
\end{align}
This decomposition removes the need to estimate $\mathcal{H}(\rvs, \rvs' | \pi)$, and instead we now only need to determine the entropy over skills $\mathcal{H}(\rvz)$. Note that, since we are free to define $p(\rvz)$ and it is independent of $\pi$, $\mathcal{H}(\rvz)$ is effectively a constant and does not influence the optimization process. Then, to obtain a tractable approximation of $\mathcal{H}(\rvz | \rvs, \rvs', \pi)$, we will introduce a variational approximation $q(\rvz | \rvs, \rvs')$ of the conditional skill distribution $p(\rvz | \rvs, \rvs', \pi)$. Since the cross-entropy between $p$ and $q$ is an upper bound on the entropy of $p$, $\mathcal{H}(p) \leq \mathcal{H}(p, q)$, we obtain the following variational lower bound on $I(\rvs, \rvs' ; \rvz | \pi)$,
\begin{align}
I(\rvs, \rvs' ; \rvz | \pi) & = \mathcal{H}(\rvz) + \expec_{p(\rvz)} \expec_{p(\rvs, \rvs' | \pi, \rvz)} \left[ \mathrm{log} \ p(\rvz | \rvs, \rvs', \pi) \right] \\
& \geq \mathop{\mathrm{max}}_q \ \mathcal{H}(\rvz) + \expec_{p(\rvz)} \expec_{p(\rvs, \rvs' | \pi, \rvz)} \left[ \mathrm{log} \ q(\rvz | \rvs, \rvs') \right] ,
\end{align}
where the lower bound is tight if $q = p$. We will refer to $q(\rvz | \rvs, \rvs')$ as the \emph{encoder}. This skill discovery objective encourages a policy to produce distinct behaviors for each skill $\rvz$, such that the encoder can easily recover the $\rvz$ that produced a particular behavior.

\subsection{Surrogate Objective}
Using the above variational approximations, we can construct a surrogate objective that approximates Equation~\ref{eqn:sd_objective},
\begin{align}
    \mathop{\mathrm{arg \ max}}_\pi \quad \expec_{p(\rvz)} \expec_{p(\tau | \pi, \rvz)} \bigg[\sum_{t=0}^{T-1} \gamma^t \big( & - \mathrm{log}\left(1 - D\left(\rvs_t, \rvs_{t+1} \right) \right)  \nonumber \\
    & + \beta \ \mathrm{log} \ q\left(\rvz | \rvs_t, \rvs_{t+1}\right) \big) \bigg] .
    \label{eqn:sur_sd_objective}
\end{align}
The reward for the policy at each timestep is then specified by
\begin{equation}
    r_t = - \mathrm{log} \left(1 - D(\rvs_t, \rvs_{t+1})\right) + \beta \ \mathrm{log} \ q\left(\rvz_t | \rvs_t, \rvs_{t+1}\right).
    \label{eqn:reward_sssd}
\end{equation}
This objective in effect encourages a model to develop a set of distinct skills, which also produce behaviors that resemble the dataset. A similar objective was previously proposed by \citet{infogan2016}
%%SL.1.24: same comment as before: when you say something is similar to a prior paper, skeptical reviewers will read it as "identical" (if they don't know that paper) unless you explain *precisely* the major difference (e.g., didn't do human motion, no animation, no RL, etc)
%%JP: added explicit mention that the infogan stuff is for image synthesis
for learning disentangled representations for image synthesis with adversarial generative networks. Similar techniques have also been applied to acquire disentangled skill representations for imitation learning \cite{infogail2017,HausmanMultiModal2017}. However, these methods have generally only been effective for low-dimensional systems, such as driving and simple RL benchmarks,
%%SL.1.9: can you give an example of what "low dimensional" means here?
%%JP: added driving as an example
%%SL.1.24: I don't think Hausman did driving? maybe just say RL benchmark tasks?
%%JP: the driving example is from the infogail paper
and have not shown to be effective for more complex domains. Furthermore, the skill embeddings learned by these prior methods have not been demonstrated as being effective for hierarchical control. In this work, we show that our method can in fact learn a rich repertoire of sophisticated motor skills for complex physically simulated characters, and in the following sections, we propose a number of design decisions that improve the quality of the resulting motions, as well as allow the skills to be effectively utilized for hierarchical control. 
\section{Low-Level Policy}

%%SL.1.9: Generally this is reasonable, but I think the opening to this section will be much more compelling if we can clearly state the problem. Something like although the method described in Section 5 conceptually provides a reasonable approach to [whatever], in practice [all these issues happen]
%%JP: reworded, hope it's more clear now
Although the method described in Section~\ref{sec:ASE} provides a conceptually elegant approach for learning skill embeddings, a number of challenges need to be addressed in order to learn effective skill representations from large datasets in practice. In this section, we detail design improvements for training an effective low-level policy $\pi(\rva | \rvs, \rvz)$, including techniques for preventing low-quality out-of-distribution samples from the latent space, improving stability during training, improving the responsiveness of the low-level policy, and developing robust recovery strategies that can be seamlessly integrated into downstream tasks.

%%SL.1.1: Remember to clearly motivate what the problem is and why it needs solving -- right now it's not quite clear why we need to worry about this part. Also, consider how to organize these subsections so that there is a clear progression, right now the subsections in Sec 6 feel a little like a laundry list, without a very clear reason for the chosen ordering
\subsection{Latent Space}
First, we consider the design of the latent space of skills $\mathcal{Z}$ and the prior over skills $p(\rvz)$. In GAN frameworks, a common choice is to model the latent distribution using a Gaussian $p(\rvz) = \mathcal{N}\left(0, I \right)$. However, this design results in an unbounded latent space, where latents that are far from the origin of $\mathcal{Z}$ may produce low-quality samples. Since the latent space will be used as the action space for a high-level policy $\omega(\rvz | \rvs, \rvg)$, an unbounded $\mathcal{Z}$ can lead $\omega$ to select latents that are far from the origin, which may result in unnatural motions. To better ensure high-quality samples, bounded latent spaces have also been used, where $p(\rvz)$ can be modeled with a uniform distribution $\mathcal{U}[-1, 1]$ or a truncated Gaussian \cite{brock2018large}. In this work, we will model the latent space as a hypersphere
%%SL.1.9: This is an intriguing decision, but the above motivation doesn't really motivate it -- in fact, up until the sentence "In this work" most readers will assume you'll use a bounded space following brock2018, making it a bit hard to understand the logic behind this interesting (but seemingly more complex) decision
$\mathcal{Z} = \{\rvz : ||\rvz|| = 1 \}$ \cite{styleGan2019}, and $p(\rvz)$ will be a uniform distribution on the surface of the sphere. Samples can be drawn from $p(\rvz)$ by normalizing samples from a standard Gaussian distribution,
\begin{equation}
    \bar{\rvz} \sim \mathcal{N}\left(0, I \right), \qquad \rvz = \bar{\rvz} / ||\bar{\rvz}|| .
\end{equation}
This provides the model with a bounded latent space, which can reduce the likelihood of unnatural behaviors arising from out-of-distribution latents. As we will discuss in Section~\ref{sec:highLevelPolicy}, this choice of latent space can also facilitate exploration for downstream tasks.

\subsection{Skill Encoder}
Since the the latent space is modeled as a hypersphere, the skill encoder will be modeled using a von Mises-Fisher distribution,
\begin{equation}
    q(\rvz | \rvs, \rvs') = \frac{1}{Z} \mathrm{exp}\left(\kappa \ \mu_q(\rvs, \rvs')^T \rvz \right) ,
\end{equation}
which is the analogue to the Gaussian distribution on the surface of a sphere. $\mu_q(\rvs, \rvs')$ is the mean of the distribution, which must be normalized $||\mu_q(\rvs, \rvs')|| = 1$, $Z$ is a normalization constant, and $\kappa$ is a scaling factor. The encoder can then be trained by maximizing the log-likelihood of samples $(\rvz, \rvs, \rvs')$ collected from the policy,
\begin{equation}
    \mathop{\mathrm{max}}_q \quad \expec_{p(\rvz)} \expec_{d^\pi(\rvs, \rvs' | \rvz)} \left[ \kappa \ \mu_q(\rvs, \rvs')^T z\right] ,
    \label{eqn:enc_objective}
\end{equation}
where $d^\pi(\rvs, \rvs' | \rvz)$ represents the likelihood of observing a state transition under $\pi$ given a particular skill $\rvz$.

\subsection{Discriminator}
Adversarial imitation learning is known to be notoriously unstable. To improve training stability and quality of the resulting motions, we incorporate the gradient penalty regularizers used by \citet{2021-TOG-AMP}. The discriminator is trained using the following objective,
\begin{align}
    \mathop{\mathrm{min}}_D \quad & -\expec_{d^\mathcal{M}(\rvs, \rvs')} \left[ \mathrm{log}\left( D\left(\rvs, \rvs' \right) \right) \right] - \expec_{d^\pi(\rvs, \rvs')} \left[ \mathrm{log}\left(1 - D\left(\rvs, \rvs' \right) \right) \right] \nonumber \\
    & + w_\mathrm{gp} \ \expec_{d^\mathcal{M}(\rvs, \rvs')} \left[ \left| \left| \nabla_\phi D(\phi) \middle|_{\phi = (\rvs, \rvs')}\right| \right|^2 \right],
    \label{eqn:disc_objective}
\end{align}
where $w_\mathrm{gp}$ is a manually specified coefficient.

\subsection{Responsive Skills}
When reusing pre-trained skills to perform new tasks, the high-level policy $\omega(\rvz | \rvs, \rvg)$
%%SL.1.1: I think this part would be easier to undestand if omega had been introduced in some overview+diagram earlier in the paper
%%JP: ok, talked about omega in the overview section and also adding an overview figure
%%SL.1.9: Maybe we can make this more explicit somehow, eg in Sec 5? Basically it feels like we need a section (could be short) that just states the obvious: what is omega, what is its objective, etc. I think this comes later, but this section is kind of conditioned on the reader understanding how that works
specifies latents $\rvz_t$ at each timestep to control the behavior of the low-level policy $\pi(\rva | \rvs, \rvz)$. A responsive low-level policy should change its behaviors according to changes in $\rvz$. However, the objective described in Equation~\ref{eqn:sur_sd_objective} can lead to unresponsive behaviors,
%%SL.1.1: If we have an ablation in the experiments section that illustrates this, it would be quite nice to forward-reference it here
%%JP: yup, will run an ablation of this and reference it here
where $\pi$ may perform different behaviors depending on the initial $\rvz_0$ selected at the start of an episode, but if a new latent $\rvz'$ is selected at a later timestep, then the policy may ignore $\rvz'$ and continue performing the same behavior specified by $\rvz_0$. This lack of responsiveness can hamper a character's ability to perform new tasks and agilely respond to unexpected perturbations.

To improve the responsiveness of the low-level policy, we propose two modifications to the objective in Equation~\ref{eqn:sur_sd_objective}. First, instead of conditioning $\pi$ on a fixed $\rvz$ over an entire episode, we will instead construct a sequence of latents $\rmZ = \{\rvz_0, \rvz_1, ..., \rvz_{T-1} \}$, and condition $\pi$ on a different latent $\rvz_t$ at each timestep $t$. The sequence of latents is constructed such that a latent $\rvz$ is repeated for multiple timesteps, before a new latent is sampled from $p(\rvz)$ and repeated for multiple subsequent timesteps. This encourages the model to learn to transition between different skills.

To further encourage the model to produce different behaviors for different latents, we incorporate a diversity objective similar to the one proposed by ~\citet{yang2018diversitysensitive} to mitigate mode-collapse of conditional GANs. These modifications lead to the following pre-training objective,
\begin{align}
    & \mathop{\mathrm{arg \ max}}_\pi \quad \expec_{p(\rmZ)} \expec_{p(\tau | \pi, \rmZ)} \bigg[ \sum_{t=0}^{T-1} \gamma^t \big( - \mathrm{log}\left(1 - D\left(\rvs_t, \rvs_{t+1} \right) \right) \nonumber \\
    & \qquad \qquad \quad \qquad \qquad \qquad \qquad \quad \ \ \ + \beta \ \mathrm{log} \ q\left(\rvz_t | \rvs_t, \rvs_{t+1}\right)  \big) \bigg] \nonumber \\
    & \quad - w_\mathrm{div} \ \expec_{d^\pi(\rvs)} \expec_{\rvz_1, \rvz_2 \sim p(\rvz)}\left[ \left(\frac{D_\mathrm{KL} \left( \pi(\cdot | \rvs, \rvz_1), \pi(\cdot | \rvs, \rvz_2) \right)}{D_\rvz \left(\rvz_1, \rvz_2\right)} - 1 \right)^2\right],
    \label{eqn:sur_sd_objective2}
\end{align}
with $w_\mathrm{div}$ being a manually specified coefficient. The last term is the diversity objective, which stipulates that if two latents $\rvz_1$ and $\rvz_2$ are similar under a distance function $D_\rvz$, then the policy should produce similar action distributions, as measured under the KL-divergence. Conversely, if $\rvz_1$ and $\rvz_2$ are different, then the resulting action distributions should also be different. In our implementation, the distance function is specified by $D_\rvz(\rvz_1, \rvz_2) = 0.5 (1 - \rvz_1^T \rvz_2)$, which reflects the cosine distance between $\rvz_1$ and $\rvz_2$, since the latents lie on a sphere. This diversity objective is reminiscent of the loss used in multidimensional scaling \cite{Kruskal1964b}, and we found it to be more stable than the objective proposed by \citet{yang2018diversitysensitive}. Note, the reward for the policy at each time step depends only on the encoder $q$ and discriminator $D$. The diversity objective is only applied during gradient updates.
%%SL.1.9: One potential issue with this section is that, by this point, the reader might start feeling kind of lost as to what the objective really is. Perhaps this issue can be mitigated somewhat by having some kind of "method summary" or "putting it all together" subsection at the end of Sec 6?

\begin{algorithm}[t!]
\caption{ASE Pre-Training}
\label{alg:pretraining}
\begin{algorithmic}[1]
\STATE{{\bf input} $\mathcal{M}$: dataset of reference motions}

\STATE{$D \leftarrow$ initialize discriminator}
\STATE{$q \leftarrow$ initialize encoder}
\STATE{$\pi \leftarrow$ initialize policy}
\STATE{$V \leftarrow$ initialize value function}

\item[]
\WHILE{not done}
    \STATE{$\mathcal{B} \leftarrow \emptyset$ \ initialize data buffer}
    \FOR{trajectory $i = 1,...,m$}
        \STATE{$\rmZ \leftarrow$ sample sequence of latents $\{\rvz_0, \rvz_1, ..., \rvz_{T-1} \}$ from $p(\rvz)$}
    	\STATE{$\tau^i \leftarrow \{\rvs_0, \rva_0, \rvs_1, ..., \rvs_T \}$ collect trajectory with $\pi$ and $\rmZ$}
        \STATE{record $\rmZ$ in $\tau^i$}
        \FOR{time step $t = 0,...,T-1$}
            \STATE{$r_t \leftarrow - \mathrm{log} \left(1 - D(\rvs_t, \rvs_{t+1})\right) + \beta \ \mathrm{log} \ q\left(\rvz_t | \rvs_t, \rvs_{t+1}\right) $}
            \STATE{record $r_t$ in $\tau^i$}
        \ENDFOR
        \STATE{store $\tau^i$ in $\mathcal{B}$}
    \ENDFOR
    
    \item[]
    \STATE{Update encoder:}
    \FOR{update step $= 1,...,n$}
        \STATE{$b^\pi \leftarrow$ sample batch of $K$ transitions $\{(\rvs_j, \rvs'_j, \rvz_j)\}_{j=1}^K$ from $\mathcal{B}$}
        \STATE{update $q$ according to Equation~\ref{eqn:enc_objective} using $b^\pi$}
    \ENDFOR
    
    \item[]
    \STATE{Update discriminator:}
    \FOR{update step $= 1,...,n$}
        \STATE{$b^{\mathcal{M}} \leftarrow$ sample batch of $K$ transitions $\{(\rvs_j, \rvs'_j)\}_{j=1}^K$ from $\mathcal{M}$}
        \STATE{$b^\pi \leftarrow$ sample batch of $K$ transitions $\{(\rvs_j, \rvs'_j)\}_{j=1}^K$ from $\mathcal{B}$}
        \STATE{update $D$ according to Equation~\ref{eqn:disc_objective} using $b^{\mathcal{M}}$ and $b^\pi$}
    \ENDFOR
    
    \item[]
    \STATE{update $V$ and $\pi$ according to Equation~\ref{eqn:sur_sd_objective2} using data from $\mathcal{B}$}
        
\ENDWHILE
\end{algorithmic}
\end{algorithm}

\subsection{Robust Recovery Strategies}
A common failure case for physically simulated characters is losing balance and falling when subjected to perturbations. In these situations, it would be favorable to have characters that can automatically recover and resume performing a task. Therefore, in addition to training the low-level policy to imitate behaviors from a dataset, $\pi$ is also trained to recover from a large variety of fallen configurations. During pre-training, the character has a $10\%$ probability of being initialized in a random fallen state at the start of each episode. The fallen states are generated by dropping the character from the air at random heights and orientations. This simple strategy then leads to robust recovery strategies that can consistently recover from significant perturbations. By incorporating these recovery strategies into the low-level controller, $\pi$ can be conveniently reused to allow the character to automatically recover from perturbations when performing downstream tasks, without requiring the character to be explicitly trained to recover from falling for each new task.

\subsection{Pre-Training}
Algorithm~\ref{alg:pretraining} provides an overview of the ASE pre-training process for the low-level policy.
%%SL.1.9: One potential misunderstanding here is that we haven't referring to this process as "pre-training" very often, so some readers might misunderstand and think that this subsection is describing a procedure for pretraining *prior* to training the GAN, when in fact it is simply using the term pre-training to refer to the GAN training.
%%SL.1.24: can you address the above? Maybe the title should be something like "Pre-Training Phase Summary" or something?
%%JP: i've added many more references to this as "pre-training" in the earlier sections. I think that should resolve some of that ambiguity?
At the start of each episode, a sequence of latents $\rmZ = \{\rvz_0, \rvz_1, ..., \rvz_{T-1}\}$ is sampled from the prior $p(\rvz)$. A trajectory $\tau^i$ is collected by conditioning the policy $\pi$ on $\rvz_t$ at each timestep $t$. The agent receives a reward $r_t$ at each timestep, calculated from the discriminator $D$ and encoder $q$ according to Equation~\ref{eqn:reward_sssd}. Once a batch of trajectories has been collected, minibatches of transitions $(\rvs_j, \rvs'_j, \rvz_j)$ are sampled from the trajectories and used to update the encoder according to Equation~\ref{eqn:enc_objective}. The discriminator is updated according to Equation~\ref{eqn:disc_objective} using minibatches of transitions $(\rvs_j, \rvs'_j)$ sampled from the agent's trajectories and the motion dataset $\mathcal{M}$. Finally, the recorded trajectories are used to update the policy. The policy is trained using proximal policy optimization (PPO) \cite{PPO2017}, with advantages computed using GAE($\lambda$) \cite{GAESchulman2015}, and the value function is updated using TD($\lambda$) \cite{Sutton1998}.

%%SL.1.1: Maybe transfer is not the right term for this? I guess you're talking about the high-level policy, maybe this section is really describing how to use [METHODNAME/MODELNAME] to address downstream tasks?
%%JP: would "High-Level Policy Training" be better?
%%SL.1.9: Yes, I think so. I think the word "design" is perhaps a bit ambiguous/confusing though.
%%JP: maybe we can just go with "high-level policy"?
\section{High-Level Policy}
\label{sec:highLevelPolicy}

After pre-training, the low-level policy $\pi(\rva | \rvs, \rvz)$ can be applied to downstream tasks by training a task-specific high-level policy $\omega(\rvz | \rvs, \rvg)$, which receives as input the state of the character $\rvs$ and a task-specific goal $\rvg$, then outputs a latent $\rvz$ for directing the low-level policy.
%%SL.1.9: Maybe just recap for the reader that the benefit of doing it this way is that the high-level doesn't have to have any motion objective or try to make things look naturalistic, because the pretrained low level takes care of all that, leaving the high level to simply focus on doing well at the task
In this section, we detail design decisions for improving exploration of skills and motion quality on downstream tasks.

\subsection{High-Level Action Space}
\label{sec:highLevelActionSpace}
\begin{figure}[t]
	\centering
    \includegraphics[width=0.95\linewidth]{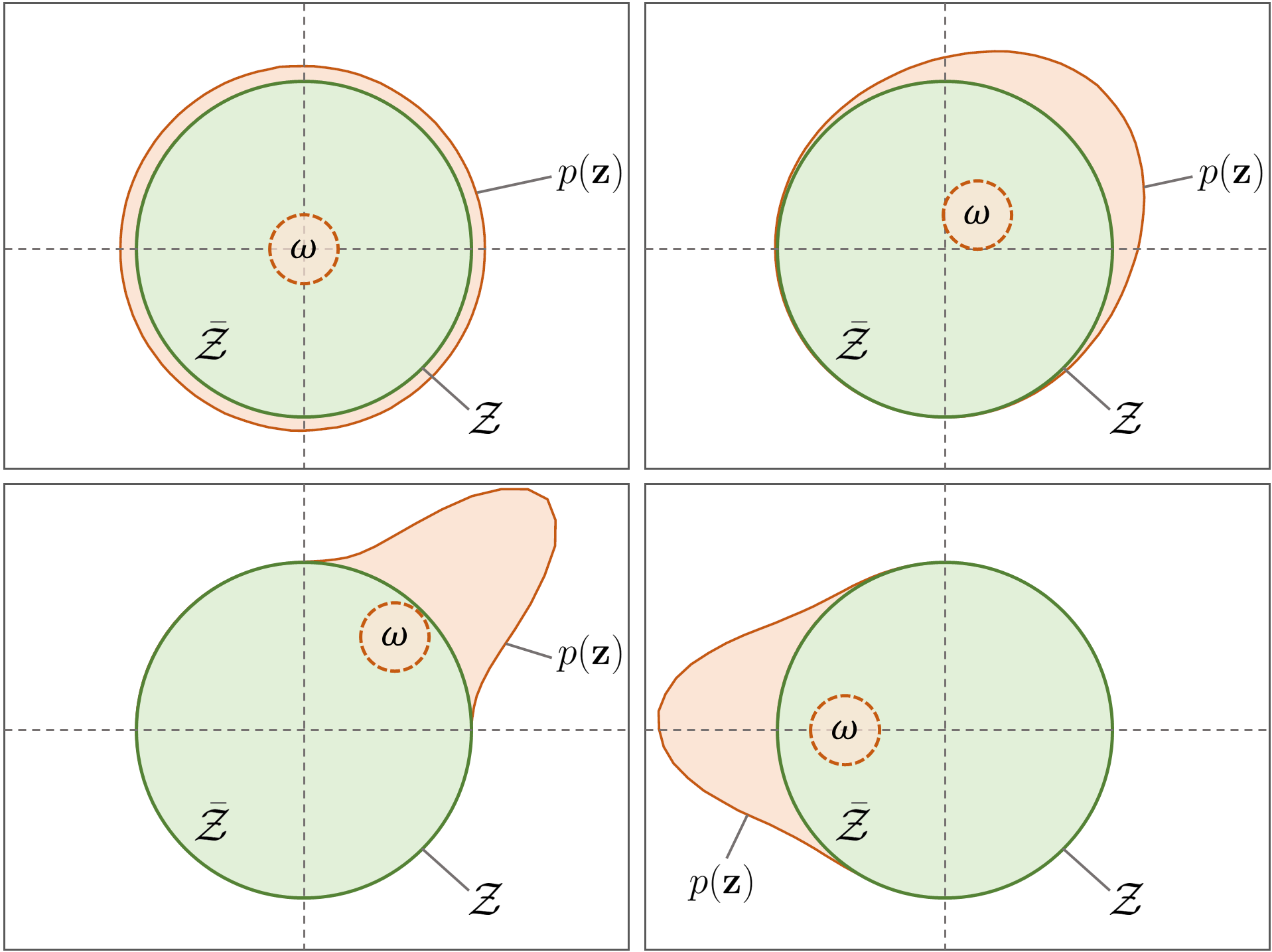}
    \caption{The \emph{unnormalized} latent space $\bar{\mathcal{Z}}$ is used as the action space for the high-level policy $\omega$. Initializing the action distribution at the origin of $\bar{\mathcal{Z}}$ allows $\omega$ to sample skills uniformly from the \emph{normalized} latent space $\mathcal{Z}$. By shifting the action distribution closer or further from the origin of $\bar{\mathcal{Z}}$, $\omega$ can increase or decrease the entropy over skills in the normalized space $\mathcal{Z}$.}
\label{fig:highLevelActionSpace}
\end{figure}

When the character is first presented with a new task, it has no knowledge of which skill will be most effective. Therefore, during early stages of training, the high-level policy should sample skills uniformly from $\mathcal{Z}$ in order to explore a diverse variety of behaviors. As training progresses, the policy should hone in on the skills that are more effective for the task, and assign lower likelihoods to skills that are less effective. Our choice of a spherical latent space provides a convenient structure that can directly encode this exploration-exploitation trade-off into the action space for the high-level policy. This can be accomplished by using the \emph{unnormalized} latents $\bar{\rvz} \in \bar{\mathcal{Z}}$ as the action space for $\omega$. The high-level policy is then defined as a Guassian in the unnormalized space $\omega(\bar{\rvz} | \rvs, \rvg) = \mathcal{N}\left(\mu_\omega(\rvs, \rvg), \Sigma_\omega \right)$. The actions from $\omega$ are projected onto $\mathcal{Z}$ by normalizing the actions $\rvz = \bar{\rvz} / ||\bar{\rvz}||$, before being passed to the low-level policy. An illustrative example of this sampling scheme is available in Figure~\ref{fig:highLevelActionSpace}.
%%SL.1.9: Cool picture!
%%JP: thanks!
At the start of training, the action distribution of $\omega$ is initialized close to the origin of $\bar{\mathcal{Z}}$, $\omega(\bar{\rvz} | \rvs, \rvg) \approx \mathcal{N}\left(0, \Sigma_\omega \right)$, which allows $\omega$ to sample skill uniformly from the normalized latent space $\mathcal{Z}$. As training progress, the mean $\mu_\omega(\rvs, \rvg)$ can be shifted away from the origin as needed by the gradient updates, thereby allowing $\omega$ to specialize, and increase the likelihood of more effective skills. By shifting the action distribution closer or further from the origin of $\bar{\mathcal{Z}}$, the $\omega$ can increase or decrease its entropy over skills in $\mathcal{Z}$ respectively.
%%SL.1.9: Generally, I found this section to be really interesting, but I think it will also be really important to include some experiment that experimentally confirms that your argument here is actually true (i.e., that this really does help with the exploration/exploitation tradeoff)
%%JP: yes, I'm planning to have some experiments to compare different choices of action space

\subsection{Motion Prior}

While $\pi$ is trained to produce natural behaviors for any latent skill $\rvz$, when reusing the low-level policy on new tasks, it is still possible for $\omega$ to specify \emph{sequences} of latents that lead to unnatural behaviors. This is especially noticeable
%%SL.1.9: perhaps "evident" is not quite the right word here?
%%JP: changed to "noticeable"
when the actions from $\omega$ changes drastically between timesteps, leading the character to constantly change the skill that it is executing, which can result in unnatural jittery movements. Motion quality on downstream tasks can be improved by reusing the discriminator $D(\rvs, \rvs')$ from pre-training as a portable motion prior when training the high-level policy. The reward for the high-level policy is then specified by a combination of a task-reward $r^G(\rvs, \rva, \rvs', \rvg)$ and a style-reward from the discriminator, in a similar manner as \citet{2021-TOG-AMP},
\begin{equation}
    r_t = w_G \ r^G(\rvs_t, \rva_t, \rvs_{t+1}, \rvg) - w_S \ \mathrm{log} \left(1 - D(\rvs_t, \rvs_{t+1}) \right) ,
\end{equation}
with $w_G$ and $w_S$ being manually specified coefficients. Note that the parameters of the discriminator are fixed after pre-training, and are not updated during task-training. Therefore, no motion data is needed
%%SL.1.9: perhaps more importantly, it doesn't need to retained or loaded in
%%JP: reworded
when training on new tasks. Prior works have observed that training a policy against a fixed discriminator often leads to unnatural behaviors that exploit idiosyncrasies of the discriminator \cite{2021-TOG-AMP}. However, we found that the low-level policy $\pi$ sufficiently constrains the behaviors that can be produced by the character, such that this kind of exploitation is largely eliminated.

\begin{figure}[t]
	\centering
    \subfigure[Simulation Model]{\includegraphics[width=0.45\columnwidth]{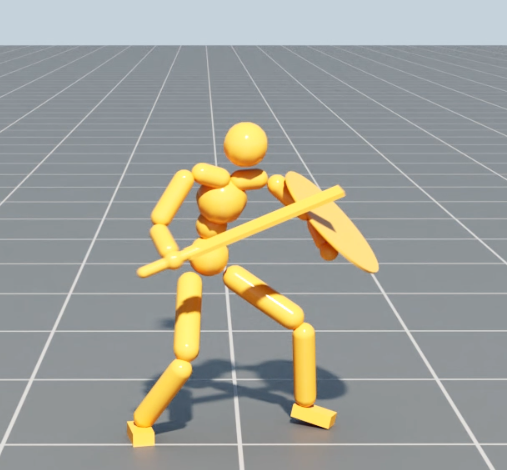}}
    \hspace{0.1cm}
    \subfigure[Visualization Model]{\includegraphics[width=0.45\columnwidth]{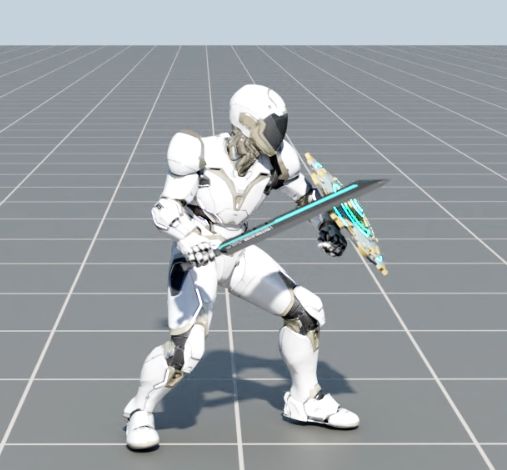}}\\
    \caption{Our framework is used to learn skill embeddings for a 37 degrees-of-freedom humanoid character, equipped with a sword and shield.}
    \label{fig:char}
\end{figure}

\section{Model Representation}
To evaluate the effectiveness of our framework, we apply ASE to develop reusable motor skills for a complex 3D simulated humanoid character, with 37 degrees-of-freedom. An illustration of the character is available in Figure~\ref{fig:char}. The character is similar to the one used by \citet{2021-TOG-AMP}, but our character is additionally equipped with a sword and shield. The sword is attached to its right hand via a 3D spherical joint, and the shield is attached to its left arm with a fixed joint. In this section, we detail modeling decisions for various components of the system. 

\subsection{States and Actions}
The state $\rvs_t$ consists of a set of features that describes the configuration of the character's body. The features include:
\begin{itemize}
    \item Height of the root from the ground.
    \item Rotation of the root in the character's local coordinate frame.
    \item Linear and angular velocity of the root in the character's local coordinate frame.
    \item Local rotation of each joint.
    \item Local velocity of each joint.
    \item Positions of the hands, feet, shield, and tip of the sword in the character's local coordinate frame.
\end{itemize}
The root is designated to be character's pelvis. The character's local coordinate frame is defined with the origin located at the root, the x-axis oriented along the root link's facing direction, and the y-axis aligned with the global up vector. The 1D rotation of revolute joints are encoded using a scalar value, representing the rotation angle. The 3D rotation of the root and spherical joints are encoded using two 3D vector corresponding to the tangent $\rvu$ and normal $\rvv$ of the link's local coordinate frame expressed in the link parent's coordinate frame \cite{2021-TOG-AMP}. Combined, these features result in a 120D state space. Each action $\rva_t$ specifies target rotations for PD controllers positioned at each of the character's joints. Like \citet{2021-TOG-AMP}, the target rotation for 3D spherical joints are encoded using a 3D exponential map \cite{ExpMapGrassia1998}. These action parameters result in a 31D action space.

\begin{figure}[t]
	\centering
    \includegraphics[width=0.9\linewidth]{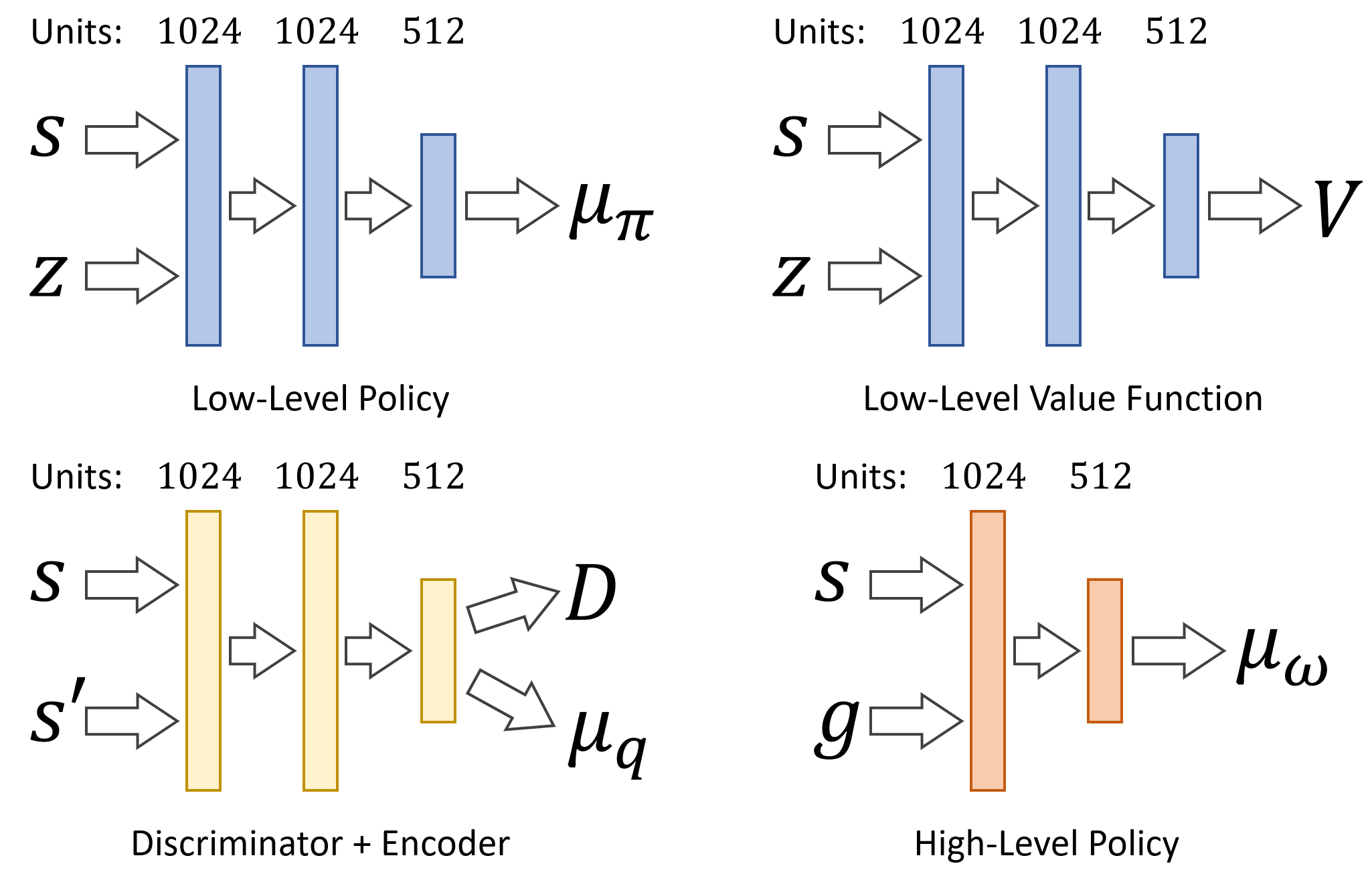}
    \caption{Network architectures used to model various  components of the system. All networks are comprised of fully-connected layers with ReLU activations for hidden layers. The discriminator $D$ and encoder $q$ are modeled by the same network with separate output layers.}
    \label{fig:nets}
\end{figure}

\begin{figure*}[t]
	\centering
    \subfigure[Sword Swing]{\includegraphics[width=1.04\columnwidth]{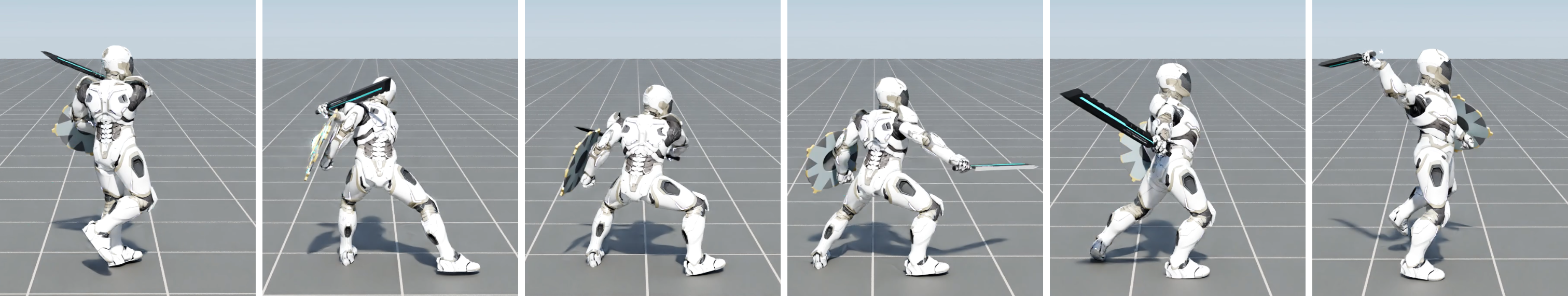}}
    \subfigure[Shield Bash]{\includegraphics[width=1.04\columnwidth]{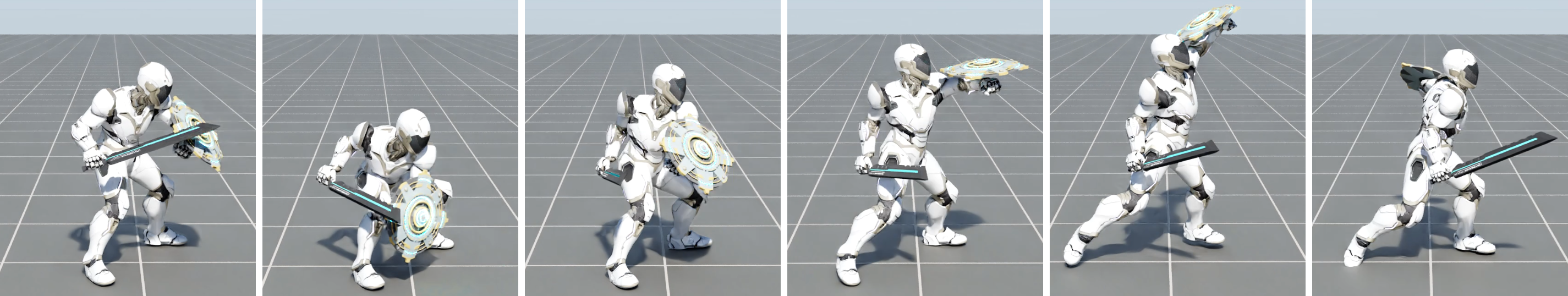}}\\
    \vspace{-0.2cm}
    \subfigure[Kick]{\includegraphics[width=1.04\columnwidth]{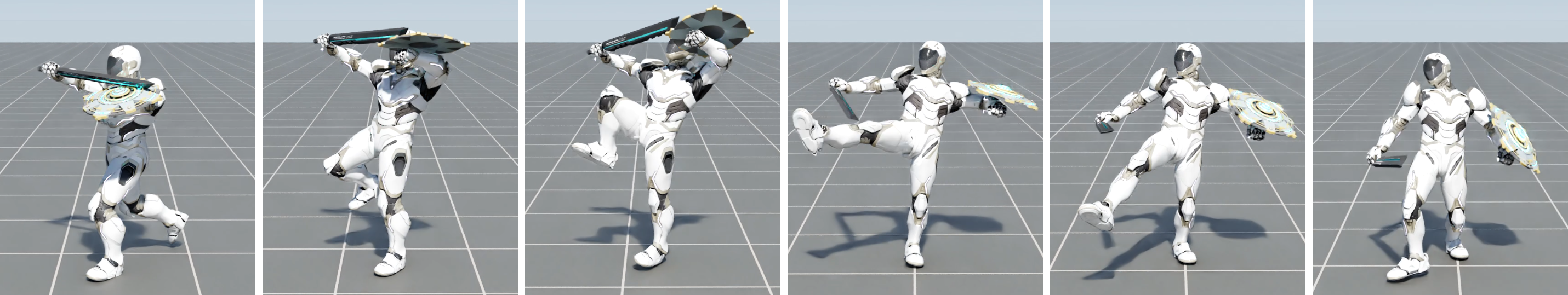}}
    \subfigure[Turn]{\includegraphics[width=1.04\columnwidth]{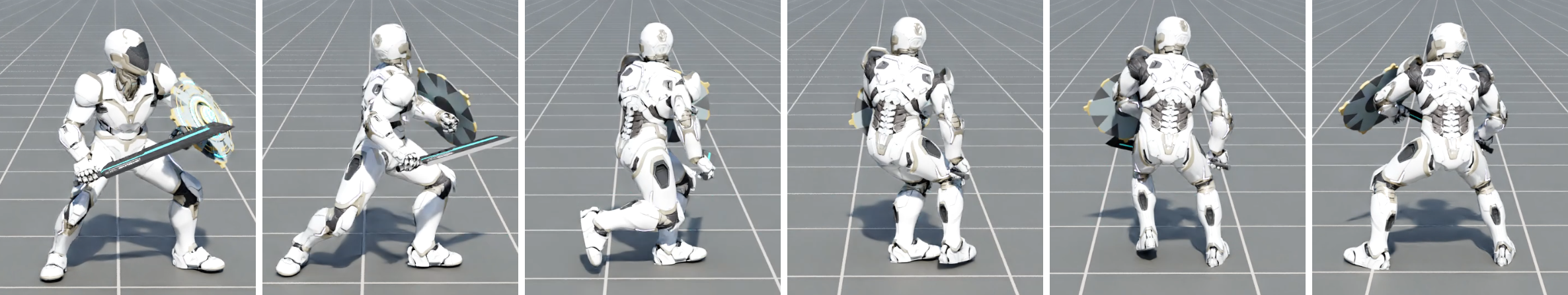}}\\
    \vspace{-0.2cm}
    \subfigure[Crouched Walk]{\includegraphics[width=1.04\columnwidth]{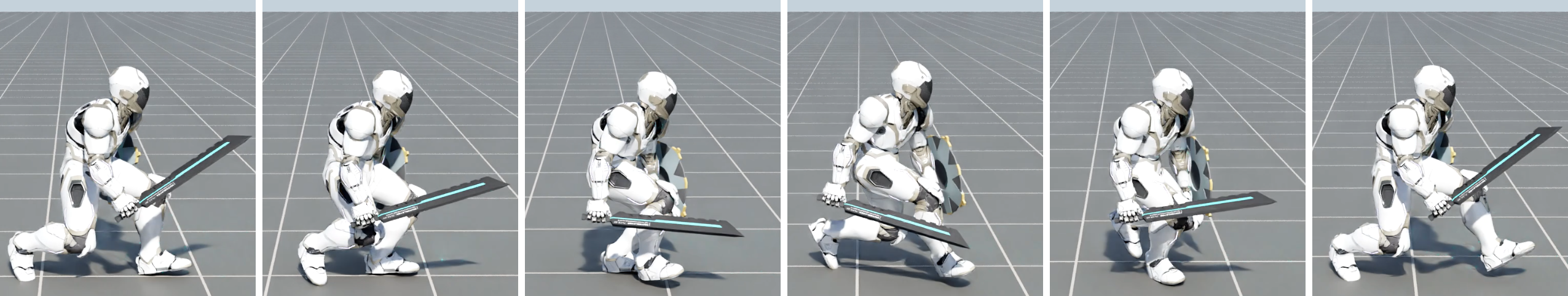}}
    \subfigure[Jog]{\includegraphics[width=1.04\columnwidth]{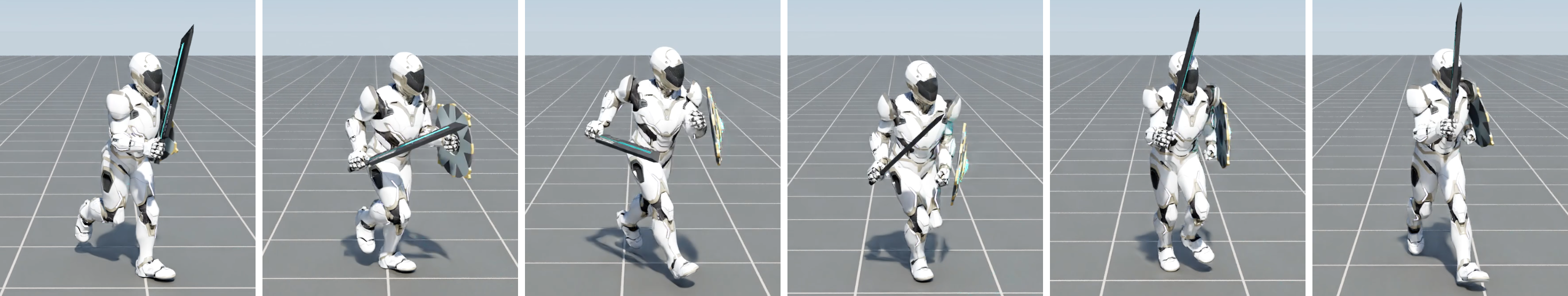}}\\
    \vspace{-0.4cm}
    \caption{Simulated character performing skills generated by random samples from the latent space. The low-level policy learns to model a diverse array of skills.}
    \label{fig:latentSnapshots}
\end{figure*}

\subsection{Network Architecture}
A schematic illustration of the network architectures used to model the various components of the system is provided in Figure~\ref{fig:nets}. The low-level policy is modeled by a neural network that maps a state $\rvs$ and latent $\rvz$ to a Gaussian distribution over actions $\pi(\rva | \rvs, \rvz) = \mathcal{N}\left(\mu_\pi(\rvs, \rvz), \Sigma_\pi \right)$, with an input-dependent mean $\mu_\pi(\rvs, \rvz)$ and a fixed diagonal covariance matrix $\Sigma_\pi$. The mean is specified by a fully-connected network with 3 hidden layers of [1024, 1024, 512] units, followed by linear output units. The value function $V(\rvs, \rvz)$ is modeled by a similar network, but with a single linear output unit. The encoder $q(\rvz | \rvs, \rvs')$ and discriminator $D(\rvs, \rvs')$ are jointly modeled by a single network, with separate output units for the mean of the encoder distribution $\mu_q(\rvs, \rvs')$ and the output of the discriminator. The output units of the encoder is normalized such that $||\mu_q(\rvs, \rvs')|| = 1$, and the output of the discriminator consists of a single sigmoid unit. The high-level policy $\omega(\bar{\rvz} | \rvs, \rvg) = \mathcal{N}\left(\mu_\omega(\rvs, \rvg), \Sigma_\omega \right)$ is modeled using 2 hidden layers with [1024, 512] units, followed by linear output units for the mean $\mu_\omega(\rvs, \rvg)$. Outputs from the high-level policy specify unnormalized latents $\bar{\rvz}$, which are then normalized $\rvz = \bar{\rvz}/||\bar{\rvz}||$ before being passed to the low-level policy $\pi$. ReLU activations are used for all hidden units \cite{ReLUNair2010}.

\section{Tasks}
Once a low-level policy has been trained to model a large variety of skills, it can then be reused to solve new downstream tasks. The corpus of tasks are designed to evaluate our model's ability to perform a diverse array of skills, compose disparate skills in furtherance of high-level task objectives, and the precision in which the model can control the character's movements. We show that the pre-trained low-level policy enable our characters to produce naturalistic motions using only simple task-reward functions.

\paragraph{Reach:}
We start with a simple reach task to evaluate the accuracy with which the model can control a character's low-level movements for tasks that require more fine-grain precision. The objective for this task is to position the tip of the sword at a target location $\rvx^*$. The goal input for the policy $\rvg_t = \tilde{\rvx}^*_t$ records the target location $\tilde{\rvx}^*_t$ in the character's local coordinate frame. The task-reward is calculated according to:
\begin{equation}
    r^G_t = \mathrm{exp} \left(-5 \ \left|\left|x^* - x^\mathrm{sword}_t \right|\right|^2 \right),
\end{equation}
where $x^\mathrm{sword}_t$ denotes the position of the sword tip at timestep $t$. The target is placed randomly within 1m of the character's root. This task represents a form of physics-based data-driven inverse-kinematics.

\paragraph{Speed:}
To evaluate the model's ability to utilize different locomotion skills, we consider a target speed task, where the objective is for the character to travel along a target direction $\rvd^*$ at a target speed $v^*$. The goal is represented by $\rvg_t = \left( \tilde{\rvd}^*_t, v^* \right)$, with $\tilde{\rvd}^*_t$ being the target direction in the character's local coordinate frame. The task-reward is calculated according to:
\begin{equation}
    r^G_t = \mathrm{exp}\left(-0.25 \left(v^* - \rvd^* \cdot \dot{\rvx}_t^\mathrm{root} \right)^2 \right),
\end{equation}
where $\dot{\rvx}_t^\mathrm{root}$ is the velocity of the character's root. The target speed is selected randomly between $v^* \in [0, 7]$m/s.

\paragraph{Steering:}
In the steering task, the objective is for the character to travel along a target direction, while facing a target heading direction $\rvh^*$. The goal is given by $\rvg_t = \left( \tilde{\rvd}^*_t, \tilde{\rvh}^*_t \right)$, with $\tilde{\rvh}^*_t$ being the local heading direction. The task-reward is calculated according to:
\begin{equation}
    r^G_t = 0.7 \ \mathrm{exp}\left(-0.25 \left(v^* - \rvd^* \cdot \dot{\rvx}_t^\mathrm{root} \right)^2 \right) + 0.3 \ \rvh^* \cdot \rvh^\mathrm{root}_t,
\end{equation}
where $\rvh^\mathrm{root}_t$ is the heading direction of the root, and the target speed is set to $v^* = 1.5$m/s.

\paragraph{Location:}
In this task, the objective is for the character to move to a target location $\rvx^*$. The goal $\rvg_t = \tilde{\rvx}_t$ records the target location $\tilde{\rvx}_t$ in the character's local frame. The task-reward is then given by:
\begin{equation}
    r^G_t = \mathrm{exp} \left(-0.5 \ \left|\left|x^* - x^\mathrm{root}_t \right|\right|^2 \right),
\end{equation}
with $x^\mathrm{root}_t$ being the location of the character's root.

\paragraph{Strike:}
Finally, to evaluate the model's effectiveness in composing disparate skills, we consider a strike task, where the objective is for the character to knock over a target object with its sword. The episode terminates if any body part makes contact with the target, other than the sword. The goal $\rvg_t = (\tilde{\rvx}^*_t, \tilde{\dot{\rvx}}^*_t, \tilde{q}^*_t, \tilde{\dot{q}}^*_t)$ records the position of the target $\tilde{\rvx}^*_t$, its rotation $\tilde{q}^*_t$, linear velocity $\tilde{\dot{\rvx}}^*_t$, and angular velocity $\tilde{\dot{q}}^*_t$. All features are recorded in the character's local coordinate frame. The reward is then calculated according to:
\begin{equation}
    r^G_t = 1 - \rvu^\mathrm{up} \cdot \rvu^*_t,
\end{equation}
where $\rvu^\mathrm{up}$ is the global up vector, and $\rvu^*_t$ is the local up vector of the target object expressed in the global coordinate frame.

\begin{figure*}[t]
	\centering
    \subfigure[Location]{\includegraphics[width=1.04\columnwidth]{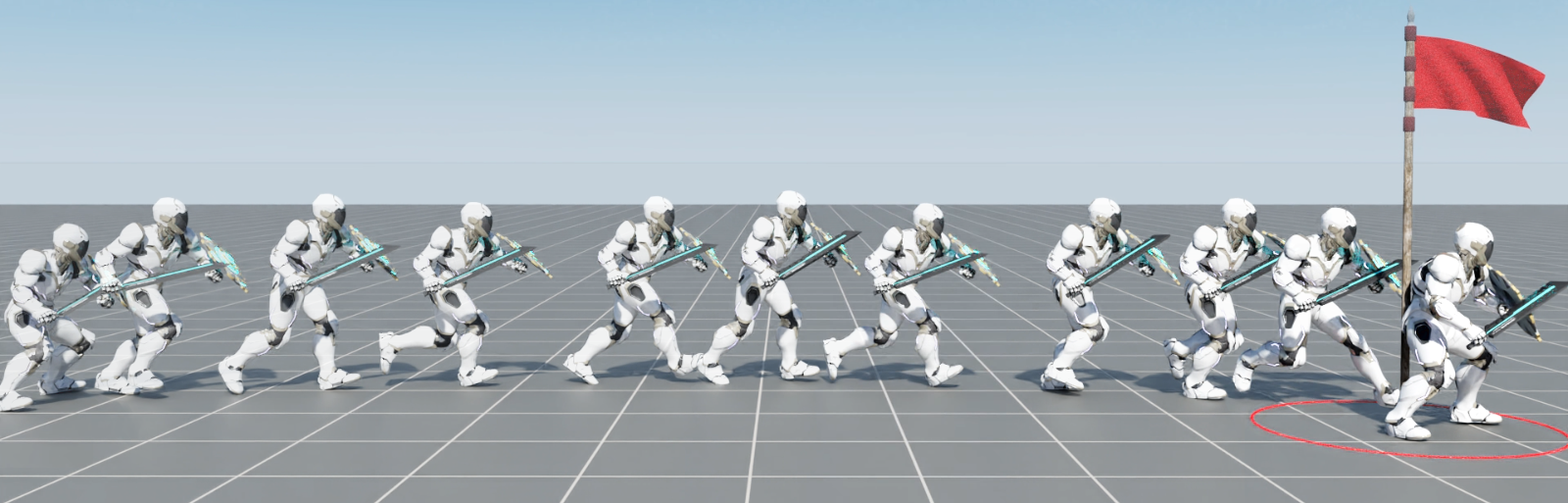}}
    \subfigure[Strike]{\includegraphics[width=1.04\columnwidth]{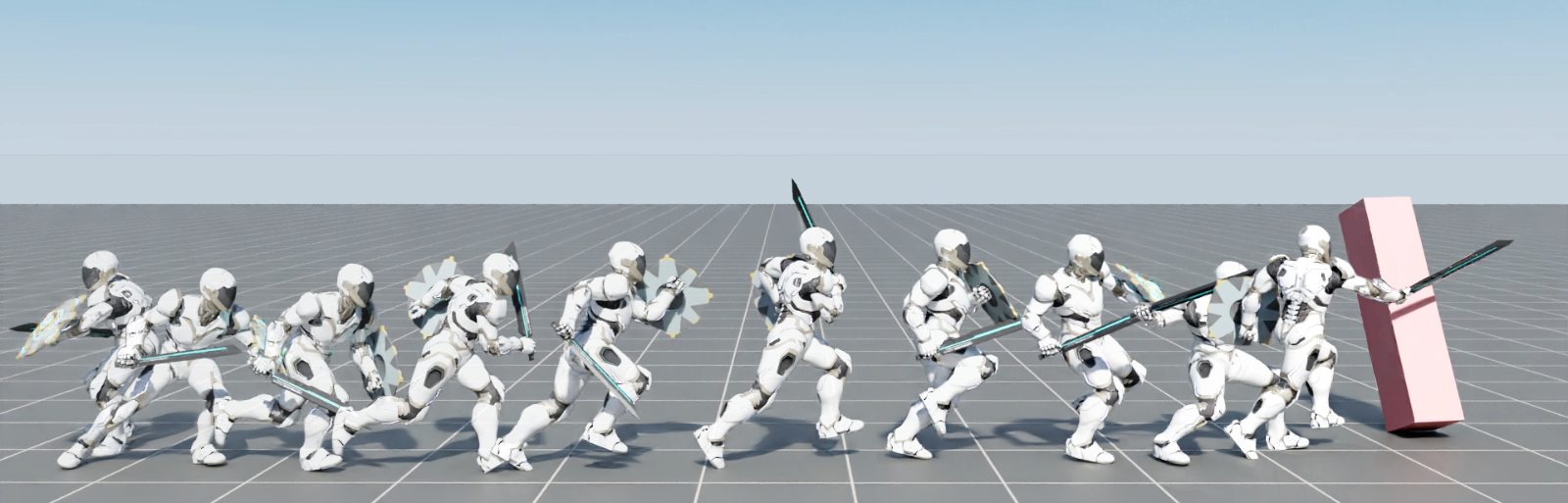}}\\
    \vspace{-0.2cm}
    \subfigure[Reach]{\includegraphics[width=1.04\columnwidth]{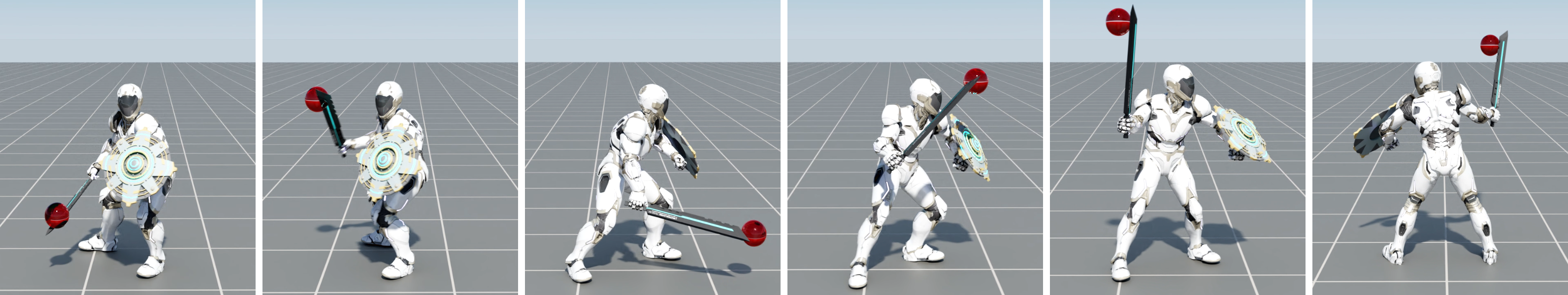}}
    \subfigure[Speed]{\includegraphics[width=1.04\columnwidth]{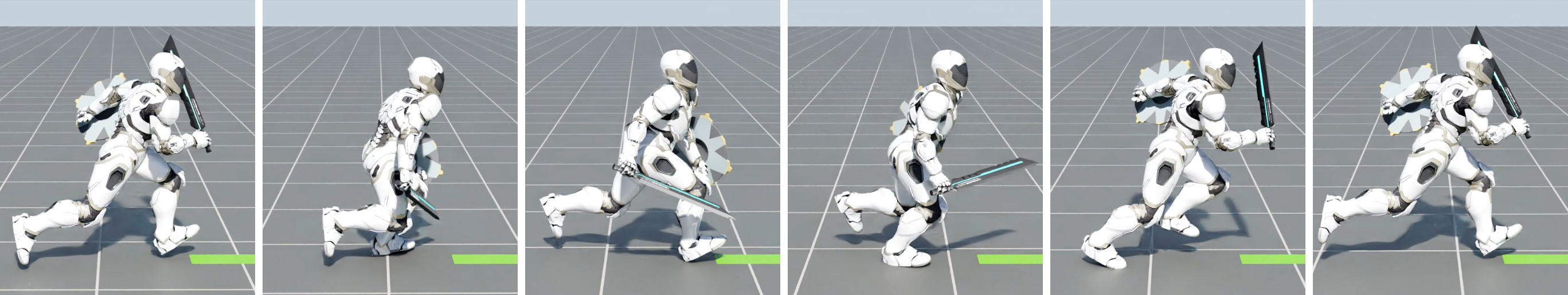}}\\
    \vspace{-0.2cm}
    \subfigure[Steering: Walking Sideways]{\includegraphics[width=1.04\columnwidth]{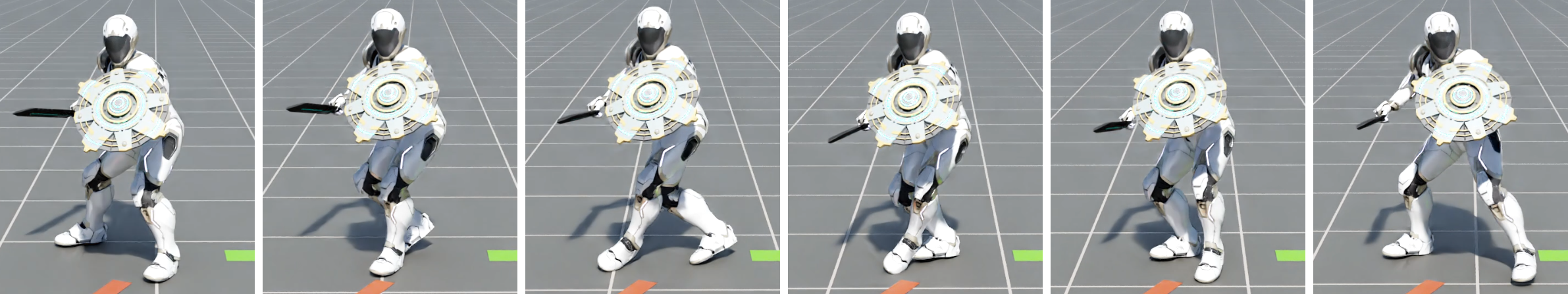}}
    \subfigure[Steering: Walking Backwards]{\includegraphics[width=1.04\columnwidth]{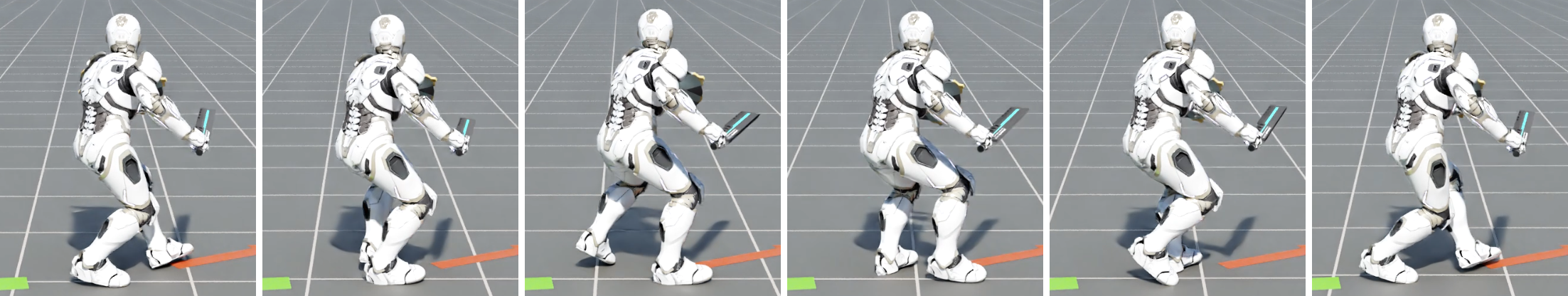}}\\
    \vspace{-0.4cm}
    \caption{Simulated character performing tasks using skills from a pre-trained low-level policy. The character can be directed to perform various tasks using simple reward functions, and the low-level policy then enables the character to achieve the task objectives by using naturalistic behaviors.}
    \label{fig:taskSnapshots}
    \vspace{-0.3cm}
\end{figure*}

\begin{figure*}[t]
	\centering
    \subfigure[Reach]{\includegraphics[width=1.04\columnwidth]{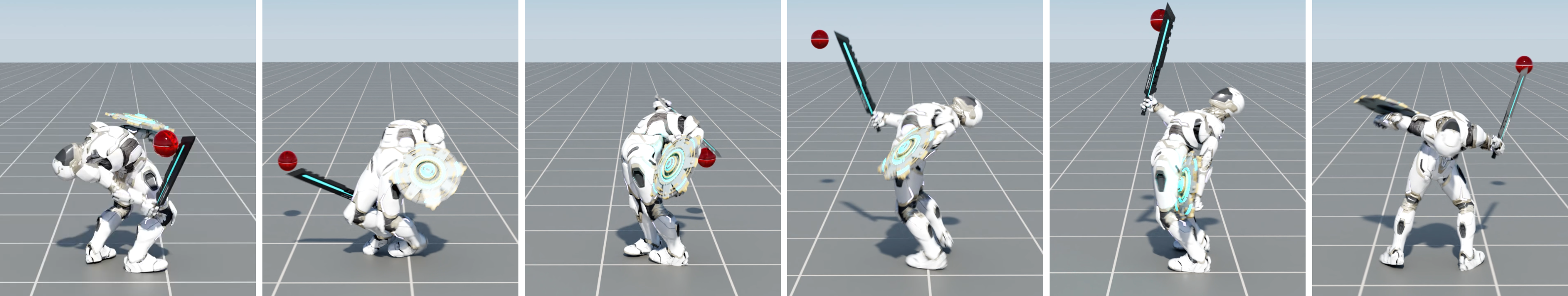}}
    \subfigure[Strike]{\includegraphics[width=1.04\columnwidth]{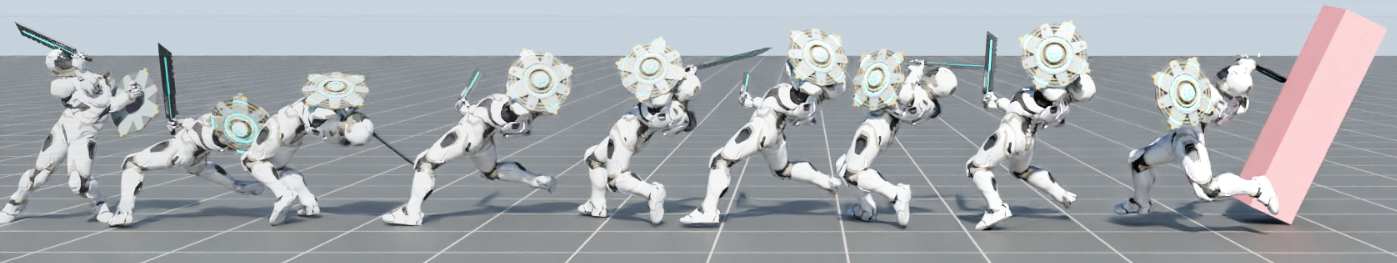}}\\
    \vspace{-0.5cm}
    \caption{Policies that are trained from scratch for each task, without using the low-level policy, often develop unnatural behaviors.}
    \label{fig:taskScratchSnapshots}
\end{figure*}

\section{Results}
We evaluate the effectiveness of our framework by using ASE to develop skill embeddings that enable a complex simulated humanoid to perform a variety of motion control tasks. First, we show that ASE can learn a rich latent embedding of diverse and sophisticated skills from large unstructured motion datasets containing over a hundred motion clips. Once trained, the learned skill model can then be reused to perform new tasks in a natural and like-like manner. Composition of disparate skills emerge automatically from the model, without requiring intricate reward shaping. Motions produced by our system are best viewed in the supplementary video.

\subsection{Experimental Setup}
%%SL.1.24: I wonder if it could make sense to have a table somewhere around here with "stats", principally focused on impressing on the reader the size of the dataset used to train the model? E.g., # of motions, # of distinct skills, etc. Not sure what exactly goes in such a table, but its goal would be to impress the reader that the model is *really big* (I think it's less important to make a big deal out of how many GPUs were used, as this in some sense is not the point -- nor really a good thing -- but how much "knowledge" from data can be packed into the model)

All environments are simulated using Isaac Gym \cite{IsaacGym2021}, a high-performance GPU-based physics simulator. During training, 4096 environments are simulated in parallel on a single NVIDIA V100 GPU, with a simulation frequency of 120Hz. The low-level policy operates at 30Hz, while the high-level policy operates at 6Hz. All neural networks are implemented using PyTorch \cite{PyTorch2019}. The low-level policy is trained using a custom motion dataset of 187 motion clips, provided by Reallusion \cite{reallusion}. The dataset contains approximately 30 minutes of motion data that depict a combination of everyday locomotion behaviors, such as walking and running, as well as motion clips that depict a gladiator wielding a sword and shield. The use of a high-performance simulator allows our models to be trained with large volumes of simulated data. The low-level policy is trained with over 10 billion samples, which is approximately 10 years in simulated time, requiring about 10 days on a single GPU. A batch of 131072 samples is collected per update iteration, and gradients are computed using mini-batches of 16384 samples. Gradient updates are performed using the Adam optimizer with a stepsize of $2 \times 10^{-5}$ \citep{AdamKingma2015}. Detailed hyperparameter settings are available in Appendix~\ref{app:hyperparams}.

\begin{figure*}[t]
	\centering
    \subfigure{\includegraphics[height=0.29\textwidth]{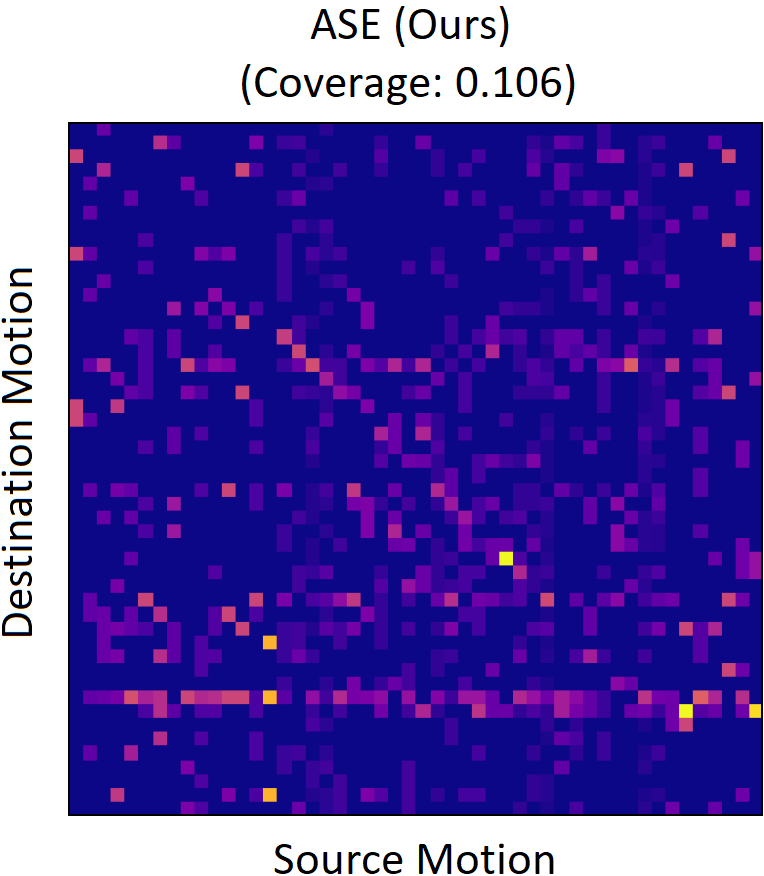}}
    \subfigure{\includegraphics[height=0.29\textwidth]{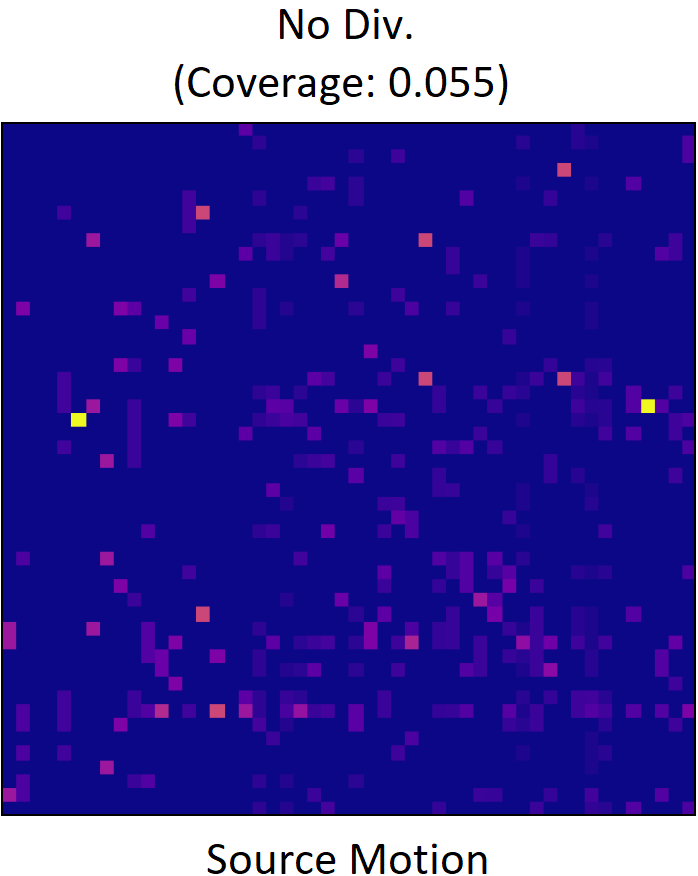}}
    \subfigure{\includegraphics[height=0.29\textwidth]{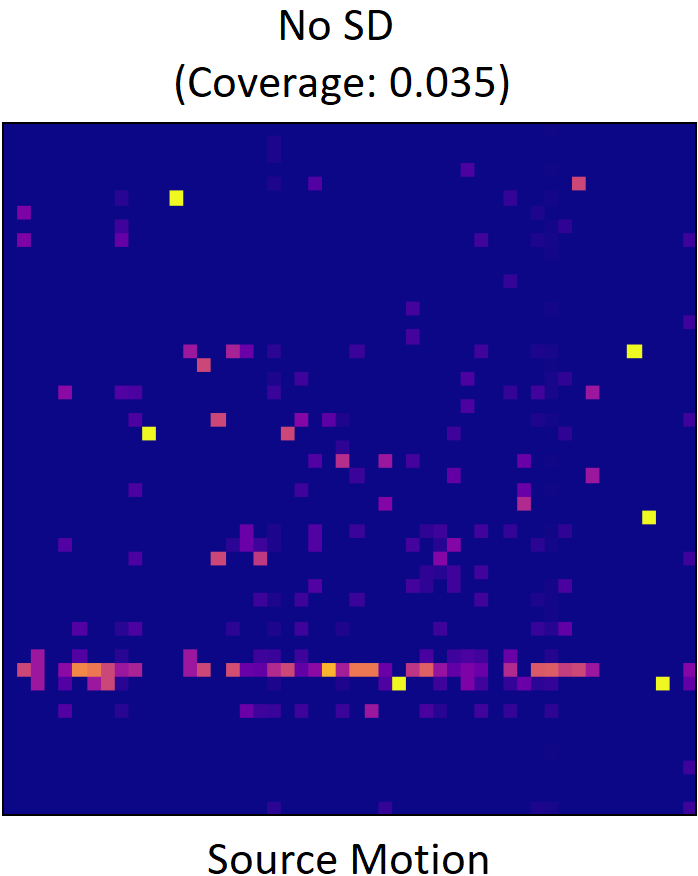}}
    \subfigure{\includegraphics[height=0.29\textwidth]{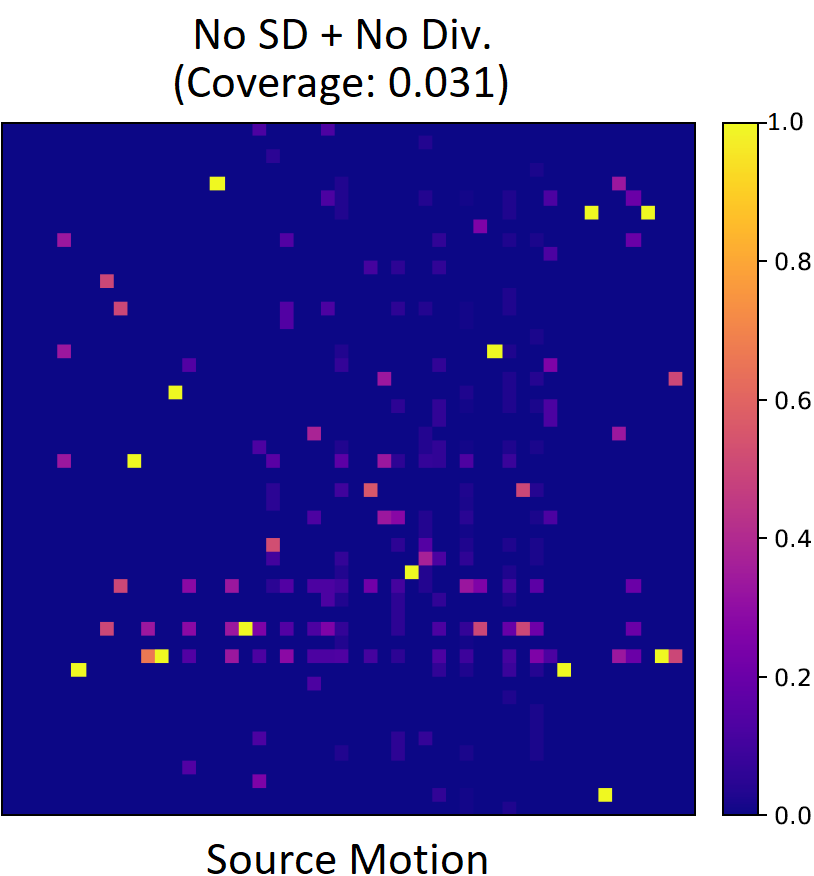}}\\
    \vspace{-0.3cm}
    %%SL.1.24: it would be good if the caption here did a better job of explaining why having dense transitions here is actually important, as this may not be obvious to all readers
    %%JP: added an extra bit to say that denser transitions can help the model compose skills for new tasks
    \caption{Probabilities of transitioning between different motion clips under various models. Each column represents the probabilities of transitioning from a source motion to each destination motion. Results are shown for the 50 most frequently matched motion clips. Coverage represents the portion of the total number of possible transitions that was observed from a model. ASE exhibits denser connections between different motions. Denser connections can lead to more responsive behaviors and improve a model's ability to compose different skills when performing new tasks. The number of transitions decrease drastically when the skill discovery objective (No SD) and the diversity objective (No Div.) are removed.}
    \label{fig:motionTransitions}
\end{figure*}

\subsection{Low-Level Skills}
The ASE pre-training process is able to develop expressive low-level policies that can perform a diverse array of complex skills. Examples of the learned skills are shown in Figure~\ref{fig:latentSnapshots}. Conditioning the policy on random latents $\rvz$ leads to a large variety of naturalistic and agile behaviors, ranging from common locomotion behaviors, such as walking and running, to highly dynamic behaviors, such as sword swings and shield bashes. All of these skills are modeled by a single low-level policy. During pre-taining, the motion dataset is treated as a homogeneous set of state transitions, without any segmentation or organization of the clips into distinct skills. Despite this lack of prescribed structure, our model learns to organize the different behaviors into a structured skill embedding, where each latent produces a semantically distinct skill, such as sword swings vs. shield bashes. This structure is likely a result of the skill discovery objective, which encourages the low-level policy to produce distinct behaviors for each latent $\rvz$, such that the encoder can more easily recover the original $\rvz$. Changing $\rvz$ partway through a trajectory also leads the character to transition to different skills. The policy is able to synthesize plausible transitions, even when the particular transitions may not be present in the original dataset.

\subsection{Tasks}
%%SL.1.24: Some of the tasks are interesting because they are impressive, while others are interesting primarily because they illustrate some ability (e.g., being very precise). But this is not really explained in this section. I think we can do a better job of explaining the goal of each task. In some sense the reach task is actually kind of lame, but it's quite important for showing that the latent space meaningfully interpolates the illustrated motions

To demonstrate the effectiveness of the pre-trained skill embeddings, we apply the pre-trained low-level policy to a variety of downstream tasks. Separate high-level policies are trained for each task, while the same low-level policy is used for all tasks. Figure~\ref{fig:taskSnapshots} illustrates behaviors learned by the character on each task. ASE learns to model a versatile corpus of skills that enables the character to effectively accomplish diverse task objectives. Though each task is represented by a simple reward function that specifies only a minimal criteria for the particular task, the learned skill embedding automatically gives rise to complex and naturalistic behaviors. In the case of the \emph{Reach} task, the reward simply stipulates that the character should move the tip of its sword to a target location. But the policy then learns to utilize various life-like full-body postures in order to reach target. The skill embedding provides the high-level policy with fine-grain control over the character's low-level movements, allowing the character to closely track the target, with an average error of $0.088 \pm 0.046$m. For the \emph{Steering} task, the character learns to utilize different forward, backward, and sideways walking behaviors to follow the target directions. When training the character to move at a target speed, the policy is able to transition between various locomotion gaits according to the desired speed. When the target speed is 0m/s, the character learns to stand still. As the target speed increases ($\sim2$m/s), the character transitions to a crouched walking gait, and then breaks into a fast upright running gait at the fastest speeds ($\sim7$m/s). This composition of disparate skills is further evident in the \emph{Strike} task, where the policy learns to utilize running behaviors to quickly move to the target. Once it is close to the target, the policy quickly transitions to a sword swing behavior in order to hit the target. After the target has been successfully knocked over, the character concludes by transitioning to an idle stance.

\begin{figure}[t]
	\centering
    \subfigure{\includegraphics[width=0.85\linewidth]{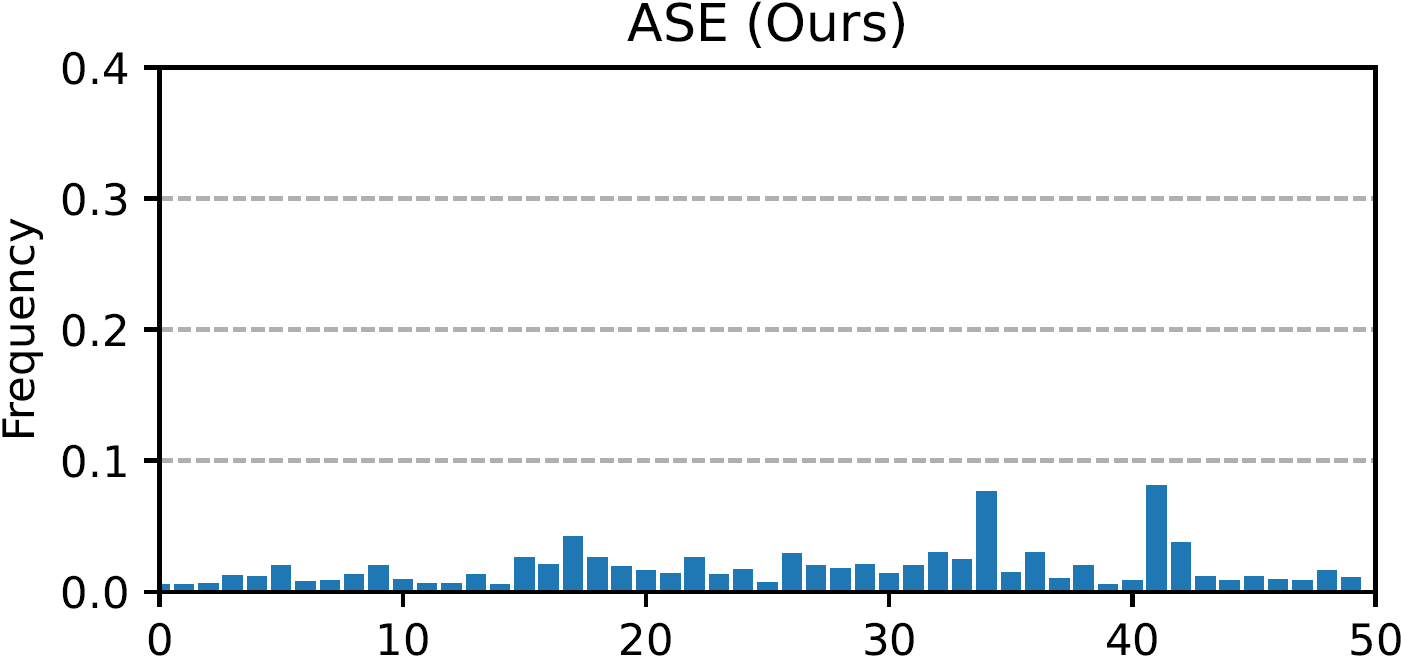}}\\
    \vspace{-0.2cm}
    \subfigure{\includegraphics[width=0.85\linewidth]{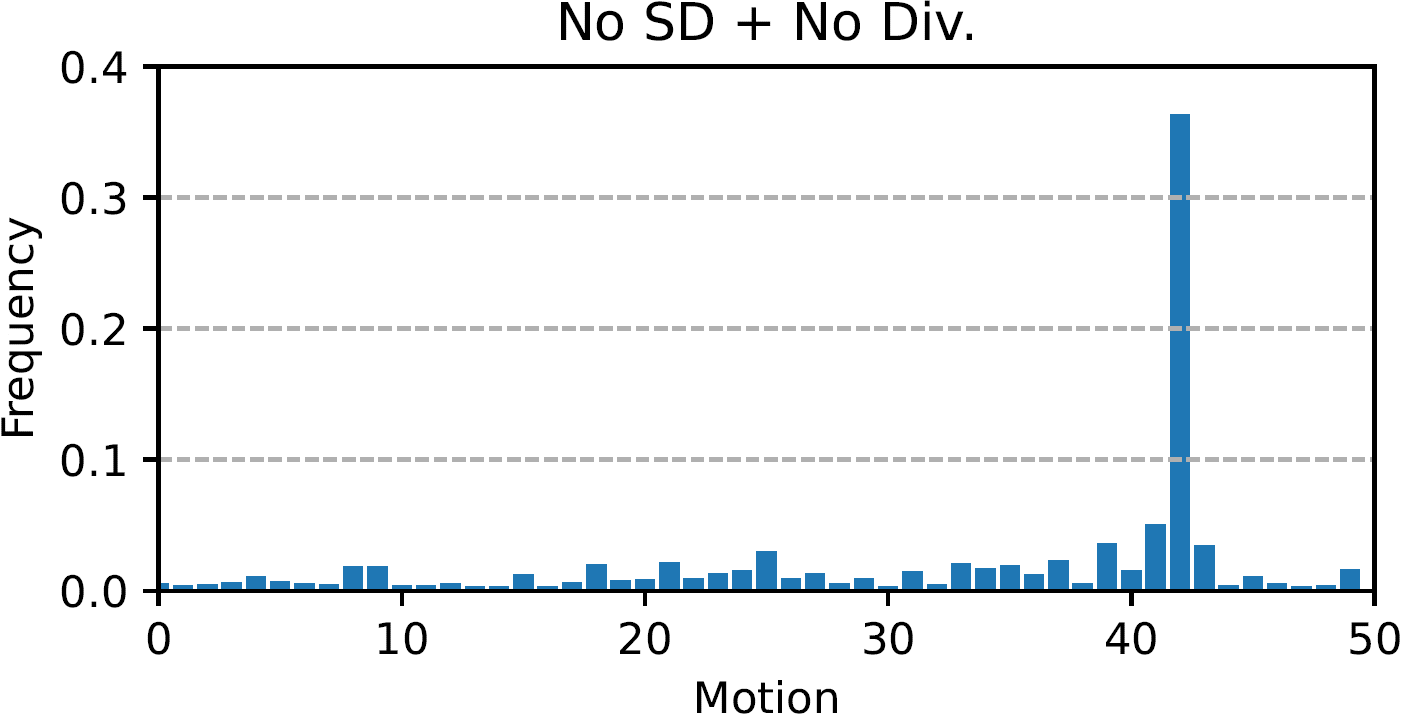}}
    \vspace{-0.4cm}
    \caption{Frequencies at which the low-level policy produces motions that match individual clips in the dataset. Results are shown for the 50 most frequently matched motion clips. ASE produces diverse behaviors that more evenly covers the dataset. Without the skill discovery objective (No SD) and the diversity objective (No Div.), the policy produces less diverse behaviors and is more prone to collapsing to a single behavior.}
    \label{fig:motionFreqs}
    \vspace{-0.4cm}
\end{figure}

\subsection{Dataset Coverage and Transitions}
Learning skill embeddings that can reproduce a wide range of behaviors is crucial for building general and reusable skill models. However, mode-collapse remains a persistent problem for adversarial imitation learning algorithms. To evaluate the diversity of the behaviors produced by our model, we compare motions produced by the low-level policy to motions from the original dataset. First, a trajectory $\tau$ is generated by conditioning $\pi$ on a random latent $\rvz \sim p(\rvz)$. Then, for every transition $(\rvs_t, \rvs_{t+1})$ in the trajectory, we find a motion clip in the dataset $m^* \in \mathcal{M}$ that contains a transition that most closely matches the particular transition from $\pi$,
\begin{equation}
    m^* = \mathop{\mathrm{arg \ min}}_{m^i \in \mathcal{M}}  \mathop{\mathrm{min}}_{(\rvs, \rvs') \in m^i} ||\bar{\rvs}_t - \bar{\rvs}||^2 + ||\bar{\rvs}_{t+1} - \bar{\rvs}'||^2 .
    \label{eqn:motionMatching}
\end{equation}
Note, $\bar{\rvs}$ represents the normalized state features of $\rvs$, normalized using the mean and standard deviation of state features from the motion data. This matching process is repeated for every transition in a trajectory, and the motion clip that contains the most matches will be specified as the clip that best matches the trajectory $\tau$. Figure~\ref{fig:motionFreqs} records the frequencies at which $\pi$ produces trajectories that matched each motion clip in the dataset across 1000 trajectories. We compare the distribution of trajectories generated by our ASE model to an ablated model that was trained without the skill discovery objective (Section~\ref{sec:skillDiscoveryObjective}) and the diversity objective (Equation~\ref{eqn:sur_sd_objective2}). The ASE model produces a diverse variety of behaviors that more evenly covers the motions in the dataset. The model trained without the skill discovery objective and the diversity object produces less diverse behaviors, where a large portion of the trajectories matched a single motion clip, corresponding to an idle standing motion.

\begin{figure}[t]
	\centering
    \subfigure{\includegraphics[height=0.35\columnwidth]{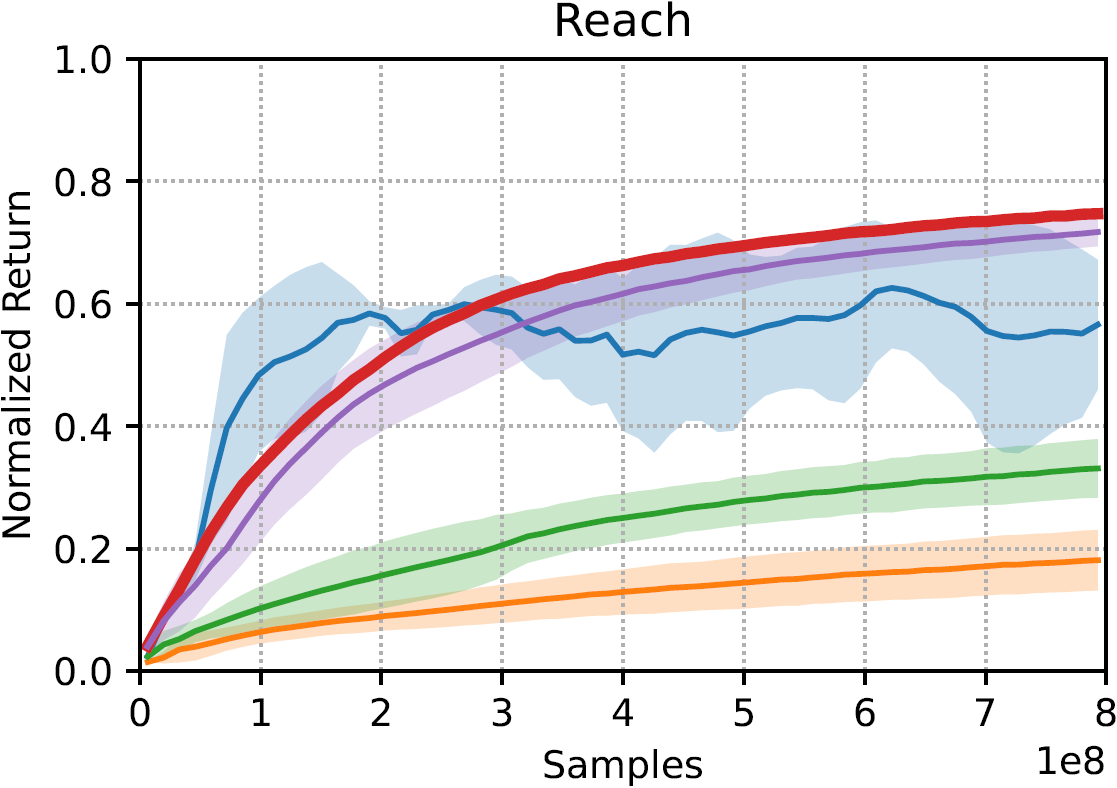}}
    \subfigure{\includegraphics[height=0.35\columnwidth]{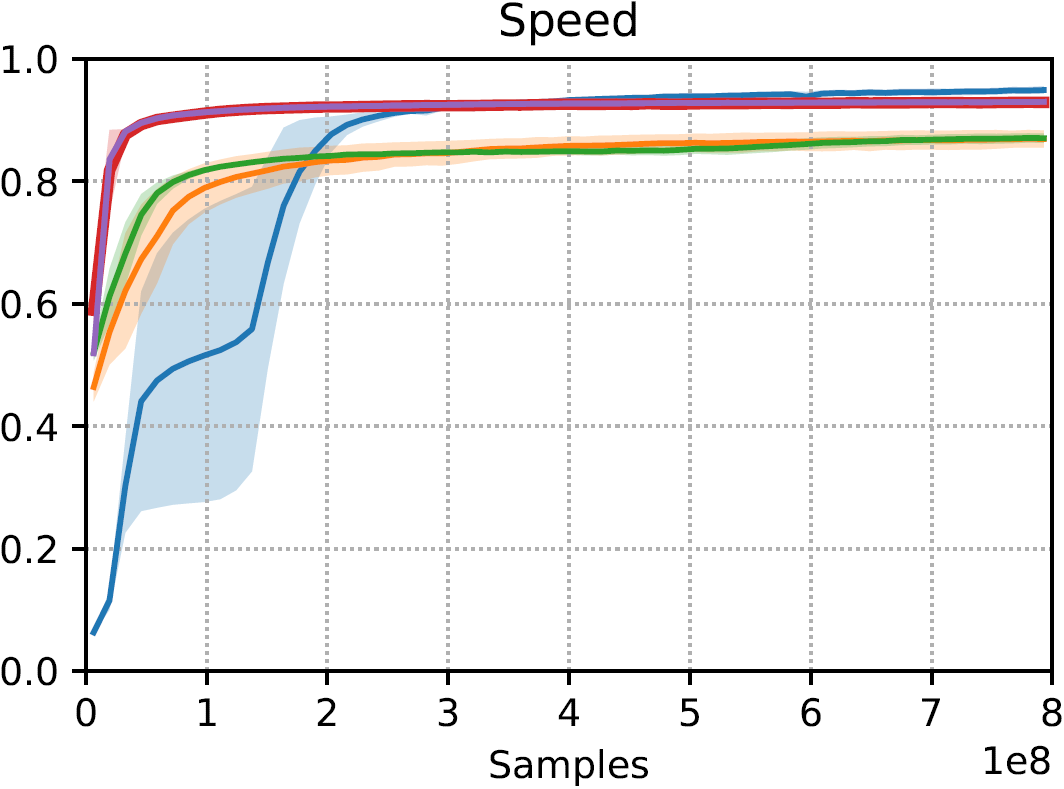}}\\
    \vspace{-0.2cm}
    \subfigure{\includegraphics[height=0.35\columnwidth]{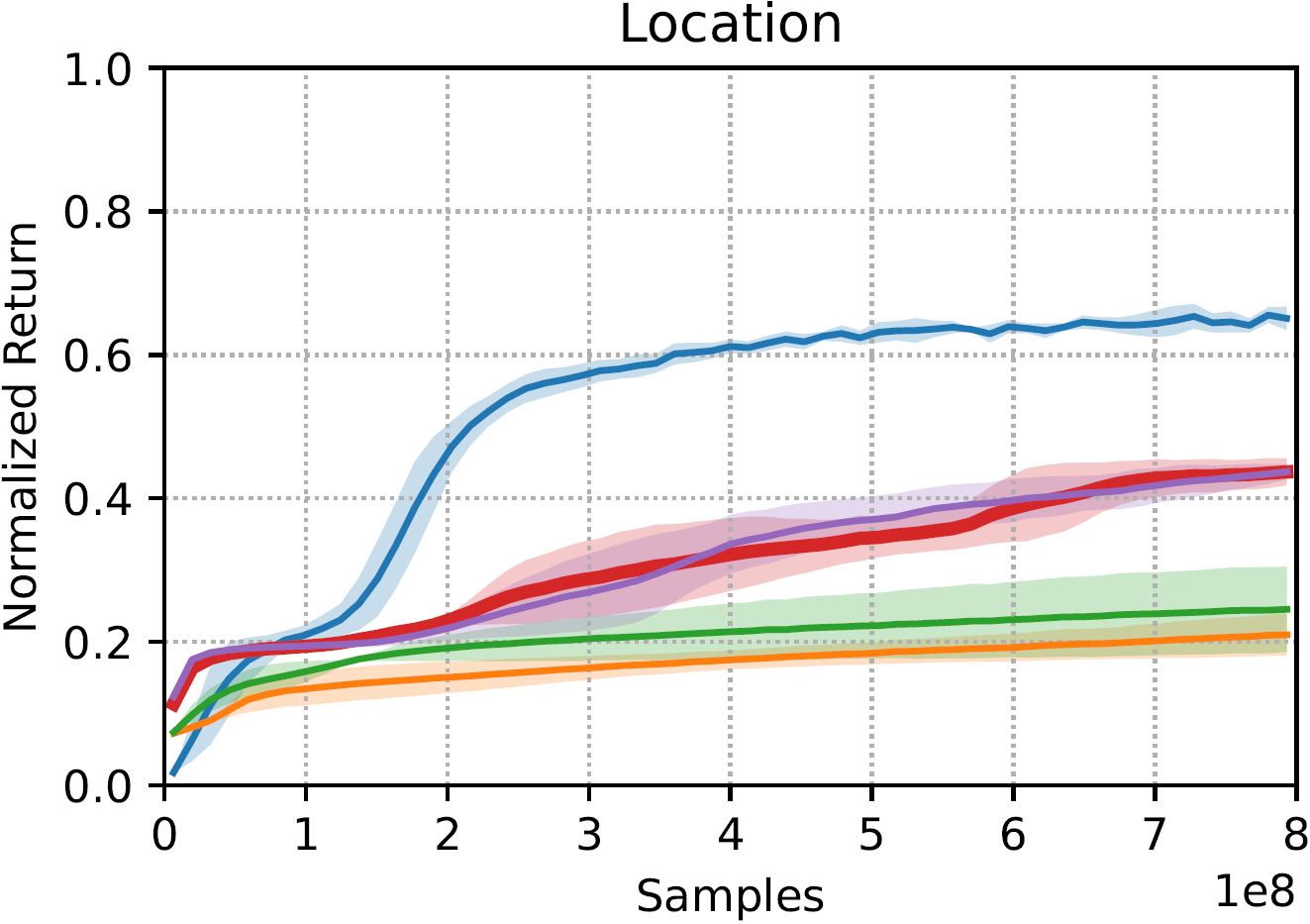}}
    \subfigure{\includegraphics[height=0.35\columnwidth]{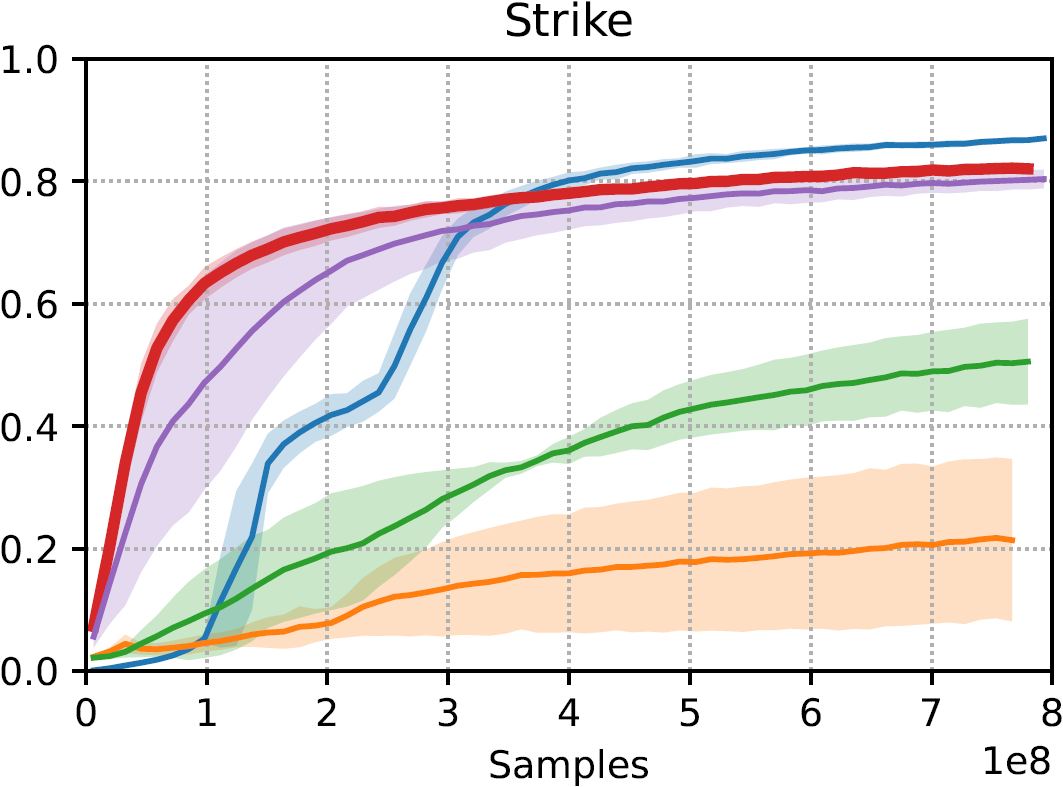}}\\
    \vspace{-0.2cm}
    \subfigure{\includegraphics[width=1\columnwidth]{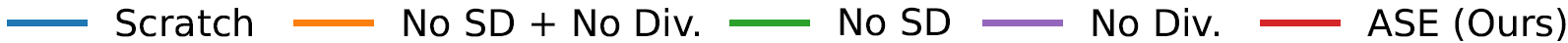}}\\
    \vspace{-0.2cm}
    %%SL.1.24: same comment as the table, it's easy for lazy readers to assume that higher y = better, but this doesn't capture motion quality... not certain what to do about this, and these graphs are less problematic than the table, but we already have tons of results, so maybe we can just remove this? or move to an appendix? just not sure if this really helps much. I guess we can also simply remove "scratch" since it doesn't actually produce even remotely reasonable motions...
    %%JP: point out that training from scratch produces terrible looking behaviors
    \caption{Learning curves comparing performance on downstream tasks using different low-level policies. We compare ASE to policies that are trained from scratch for each tasks (Scratch), as well as to low-level policies trained without the skill discovery objective (No SD), without the diversity objective (No Div.), and with both objectives disabled (No SD + No Div.). The skill discovery objective is crucial for learning effective skill representations. The policies trained from scratch often achieve higher returns by exploiting unnatural behaviors (see Figure~\ref{fig:taskScratchSnapshots})}
    \label{fig:perfCurves}
\end{figure}

In addition to producing diverse skills, the low-level policy should also learn to transition between the various skills, so that the skills can be more easily composed and sequenced to perform more complex tasks. To evaluate a model's ability to transition between different skills, we generate trajectories by conditioning $\pi$ on two random latents $\rvz_S$ and $\rvz_D$ per trajectory. A trajectory is then generated by first conditioning $\pi$ on $\rvz_S$ for 150 to 200 timesteps, then $\pi$ is conditioned on $\rvz_D$ for the remainder for the trajectory. We will refer to the two sub-trajectories produced by the different latents as the source trajectory $\tau_S$ and the destination trajectory $\tau_D$. The two sub-trajectories are separately matched to motion clips in the dataset, following the same procedure in Equation~\ref{eqn:motionMatching}, to identify the source motion $m^S$ and destination $m^D$. We repeat this process for about 1000 trajectories, and record the probability of transitioning between each pair of motion clips in Figure~\ref{fig:motionTransitions}. We compare ASE to models trained without the skill discovery objective (No SD), without the diversity objective (No Div.), and without both objectives (No SD + No Div.). Coverage denotes the portion of transitions out of all possible pairwise transitions observed from trajectories produced by a model $\left(\mathrm{coverage} = \frac{\mathrm{transitions \ from \ model}}{\mathrm{all \ possible \ transitions}}\right)$. Our ASE model generates much denser connections between different skills, and exhibits about $10\%$ of all possible transitions. Removing the skill discovery objective and diversity objective results in less responsive models that exhibit substantially fewer transitions between skills.

\begin{table}[t]
{ \centering  
%%SL.1.24: This table doesn't really make a good impression, because it makes it look like "Scratch" is the best, but of course it completely ignores how realistic the behavior is. I wonder if we should simply omit this table? It's pretty misleading... but if we do have it, it feels like we need to somehow in the same table or somewhere close by clearly show that Scratch is very unrealistic.
%%JP: explicitly mention that training from scratch looks unnatural and point to figure 7
\caption{Performance of the different skill models when applied to various tasks. Performance is recorded as the normalized return, with 0 being the minimum possible return per episode, and 1 being the maximum. The returns are averaged across 3 models using different pre-trained low-level policies, with 4096 episodes per model. The policies trained from scratch achieve higher returns on most tasks by utilizing unnatural behaviors.}
\label{tab:taskPerf}
\resizebox{\columnwidth}{!}{
\begin{tabular}{|l|c|c|c|c|c|}
\hline
{\bf Task} & {\bf Scratch} & {\bf \shortstack{No SD \\ + No Div.}} & {\bf No SD} & {\bf No Div.} & {\bf \shortstack{ASE \\ (Ours)}} \\ \hline
    Reach & $0.56 \pm 0.11$ & $0.18 \pm 0.05$ & $0.33 \pm 0.05$ & $0.72 \pm 0.02$ & $\cellcolor{gray!30} \bf{0.75 \pm 0.01}$ \\ \hline
    Speed & $\cellcolor{gray!30} \bf{0.95 \pm 0.01}$ & $0.87 \pm 0.01$ & $0.87 \pm 0.01$ & $0.93 \pm 0.01$ & $0.93 \pm 0.01$ \\ \hline
    Steering & $\cellcolor{gray!30} \bf{0.94 \pm 0.01}$ & $0.72 \pm 0.01$ & $0.74 \pm 0.02$ & $0.90 \pm 0.01$ & $0.90 \pm 0.01$ \\ \hline
    Location & $\cellcolor{gray!30} \bf{0.67 \pm 0.01}$ & $0.22 \pm 0.04$ & $0.25 \pm 0.06$ & $0.47 \pm 0.01$ & $0.45 \pm 0.01$ \\ \hline
    Strike & $\cellcolor{gray!30} \bf{0.87 \pm 0.01}$ & $0.21 \pm 0.13$ & $0.50 \pm 0.07$ & $0.80 \pm 0.02$ & $0.82 \pm 0.01$ \\ \hline
\end{tabular}
}
}
\end{table}

\begin{figure*}[t]
	\centering
    \subfigure{\includegraphics[height=0.21\textwidth]{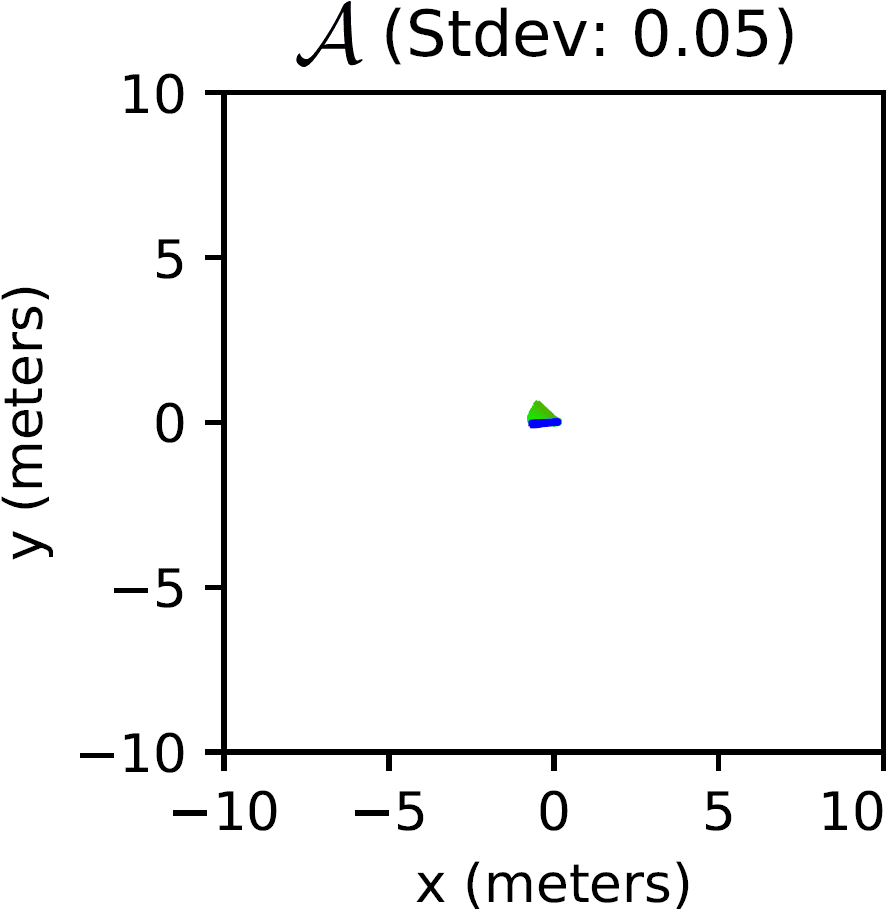}}
    \subfigure{\includegraphics[height=0.21\textwidth]{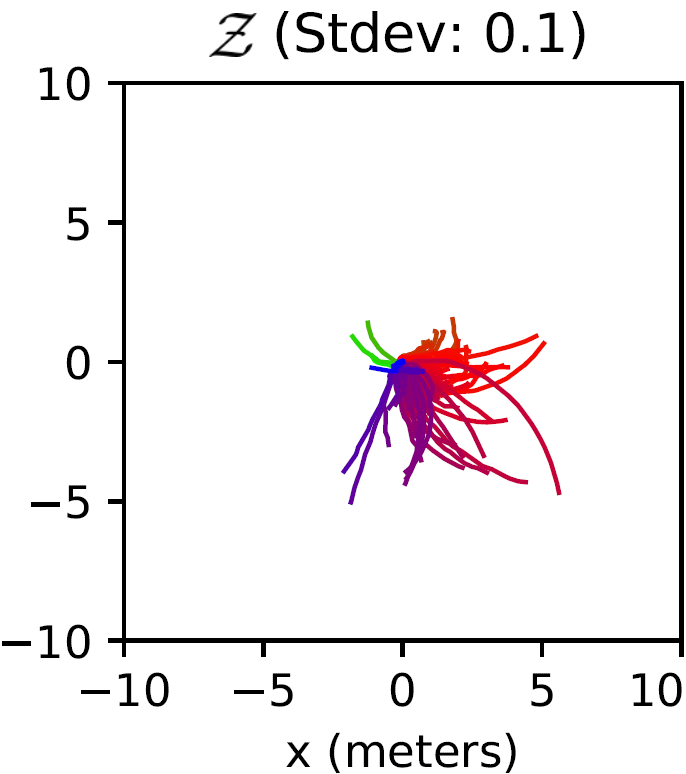}}
    \subfigure{\includegraphics[height=0.21\textwidth]{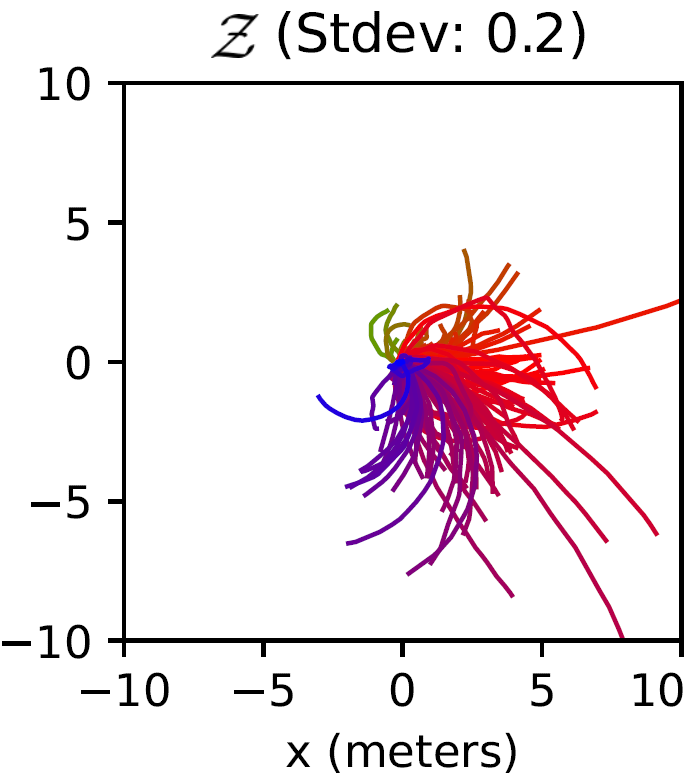}}
    \subfigure{\includegraphics[height=0.21\textwidth]{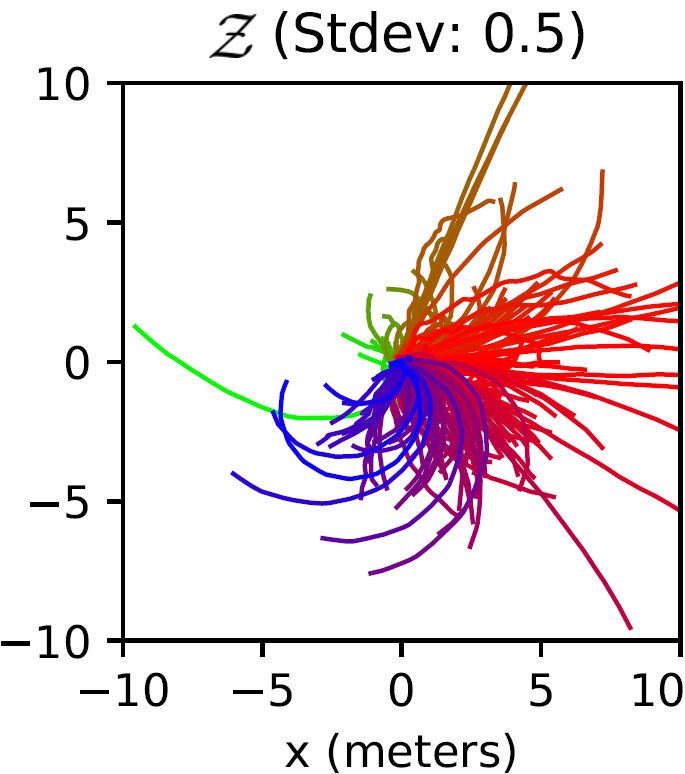}}
    \subfigure{\includegraphics[height=0.21\textwidth]{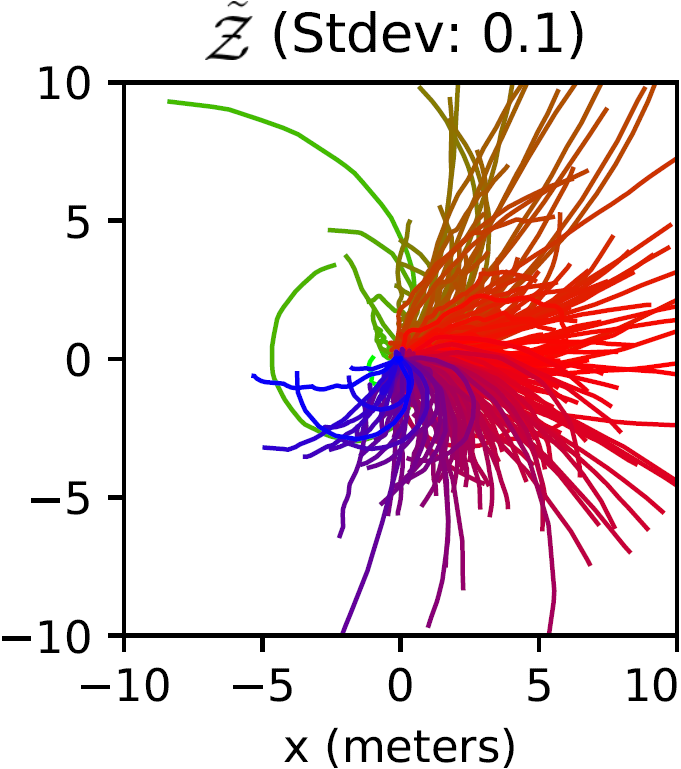}}
    \vspace{-0.3cm}
    \caption{Trajectories of the character's root produced by random exploration with different action spaces for the high-level policy. Random exploration in the original action space $\mathcal{A}$ does not produce semantically meaningful behaviors, and tends to cause the character to fall after a few timesteps. Our method of using the unnormalized latent space $\tilde{\mathcal{Z}}$ as the action space allows the policy to explore more structured and diverse behaviors. Using the normalized latent space $\mathcal{Z}$ can lead to less diverse behaviors, and a much larger standard deviation is needed for the action distribution to produce similar diversity.}
    \label{fig:highLevelTrajs}
\end{figure*}

To determine the performance impact of these design decisions, we compare the task performance achieved using various low-level policies trained with and without the skill discovery objective and diversity objective. Figure~\ref{fig:perfCurves} compares the learning curves of the different models, and Table~\ref{tab:taskPerf} summarizes the average return of each model. The skill discovery objective is crucial for learning an effective skill representation. Without this objective, the skill embedding tends to produce less diverse and distinct behaviors, which in turn leads to a significant deterioration in task performance. While the diversity objective did lead to denser connections between skills (Figure~\ref{fig:motionTransitions}), removing this objective (No Div.) did not seem to have a large impact on performance on our suite of tasks. For most of the tasks we considered, the highest returns are achieved by policies that were trained from scratch for each task. These policies often utilize unnatural behaviors in order to better maximize the reward function, such as adopting highly energetic sporadic movements to propel the character more quickly towards the target in the \emph{Location} and \emph{Strike} tasks. Examples of these behaviors are shown in  Figure~\ref{fig:taskScratchSnapshots}. Training from scratch also tends to require more samples compared policies trained using ASE, since the model needs to repeatedly relearn common skills for each task, such as maintaining balance and locomotion. ASE is able to achieve high returns on the suite of tasks, while also producing more natural and life-like behaviors.

\begin{figure}[t]
	\centering
    \subfigure{\includegraphics[width=\columnwidth]{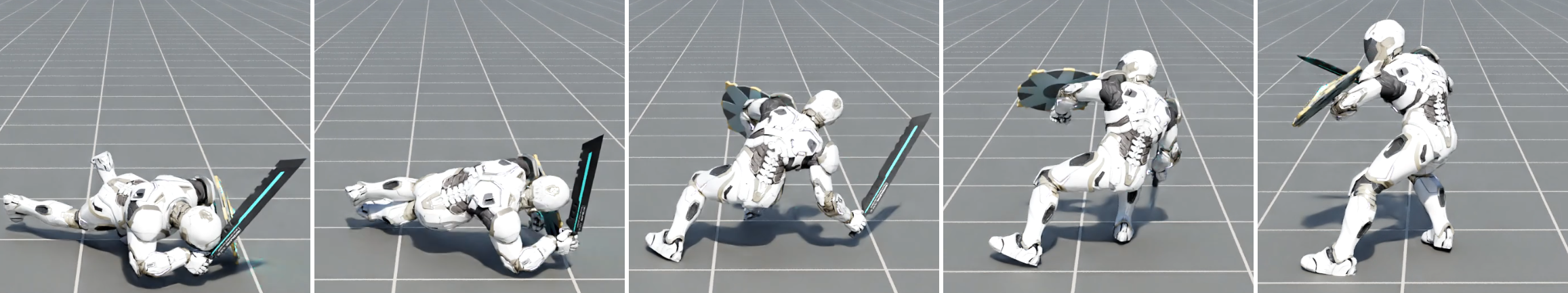}}\\
    \vspace{-0.3cm}
    \subfigure{\includegraphics[width=\columnwidth]{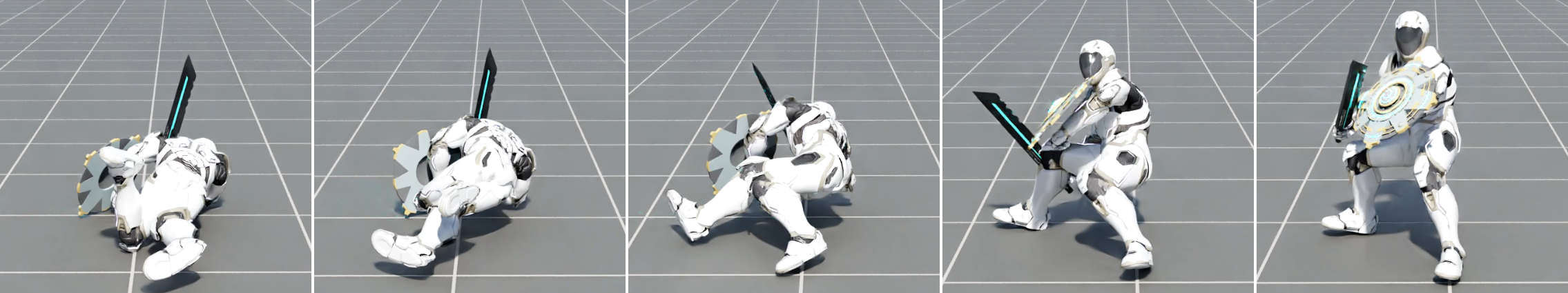}}\\
    \vspace{-0.3cm}
    \subfigure{\includegraphics[width=\columnwidth]{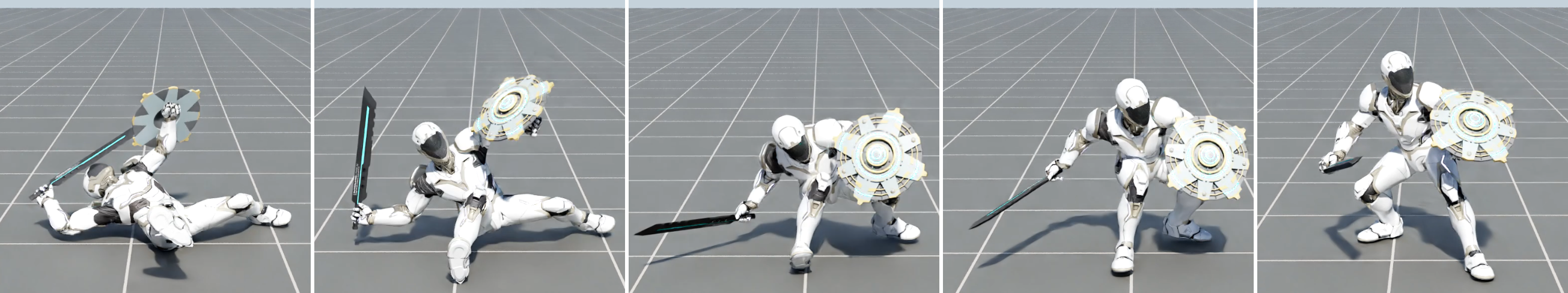}}\\
    \vspace{-0.3cm}
    \caption{The low-level policy can consistently recover after falling. The dataset does not contain motion clips that depict get up behaviors, however our model still develops plausible recovery strategies.}
    \label{fig:recoverySnapshots}
\end{figure}

\subsection{High-Level Action Space}
To evaluate the effects of different choices of action space for the high-level policy, we visualize the behaviors produced by random exploration in different action spaces. Figure~\ref{fig:highLevelTrajs} illustrates trajectories produced by the different action spaces, and the corresponding motions can be viewed in the supplementary video. First, we consider the behaviors produced by random exploration in the original action space of the system $\mathcal{A}$. This strategy does not produce semantically meaning behaviors, and often just leads to the character falling after a few timesteps. Next, we consider behaviors produced by sampling from a Gaussian distribution in the unnormalized latent space $\tilde{\mathcal{Z}}$, with a standard deviation of 0.1, as described in Section~\ref{sec:highLevelActionSpace}. Positioning the Gaussian at the origin of $\tilde{\mathcal{Z}}$ allows the policy to uniformly sample latents from the normalized latent space $\mathcal{Z}$, leading to a diverse range of behaviors that travel in different directions. Finally, we have trajectories produced by directly sampling in the normalized latent space $\mathcal{Z}$ with various standard deviations $\sigma = [0.1, 0.2, 0.5]$. This sampling strategy limits the policy to selecting latents within a local region on the surface of a hypersphere, which can lead to less diverse behaviors. More diverse behaviors can be obtained by drastically increasing the standard deviation of the action distribution. However, a large action standard deviation can hamper learning, and additional mechanisms are needed to decrease the standard deviation over time. 
%Figure~\ref{fig:actionSpacePerf} compares learning curves on downstream tasks using the different action spaces. Sampling from $\tilde{\mathcal{Z}}$ allows the policy to explore a more diverse range of behaviors at the early stages of training, while also allowing the policy specialize in more effective behaviors as training progresses. This leads faster learning and more optimal asymptotic performance compared to sampling directly from $\mathcal{Z}$.

\subsection{Motion Prior}
Reusing the pre-trained discriminator as a motion prior when training a high-level policy can improve motion quality. The task-reward function is combined with the style-reward using weights $w_G = 0.9$ and $w_S = 0.1$ respectively. A comparison of the motions produced with and without a motion prior is available in the supplementary video. Without the motion prior, the character is prone to producing unnatural jittering behaviors and other extraneous movements. With the motion prior, the character produces much smoother motions, and exhibits more natural transitions between different skills.

\begin{figure}[t]
	\centering
    \includegraphics[width=0.8\linewidth]{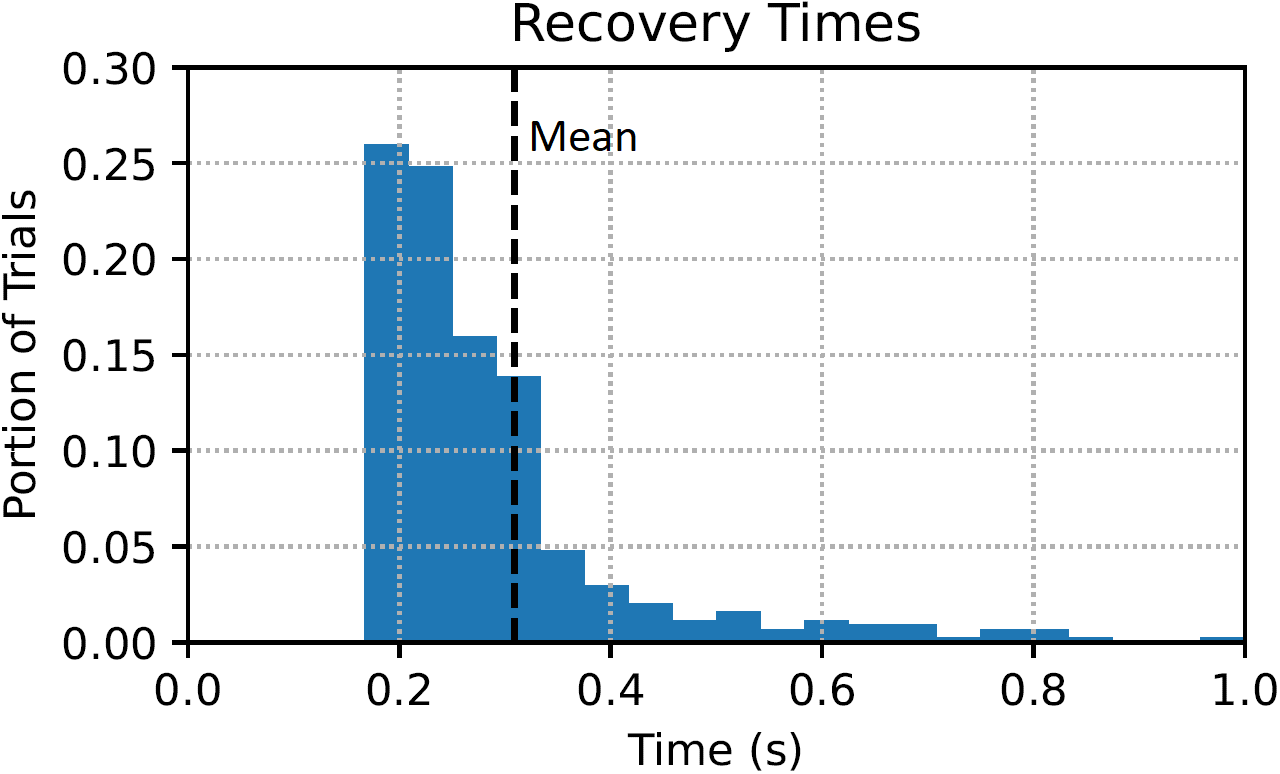}
    \vspace{-0.2cm}
    \caption{Time required for the character to get back up after falling when subjected to random perturbation forces. The low-level policy is able to consistently recover after falling across 500 trials, requiring an average recovery time of 0.31s and a maximum of 4.1s.}
    \label{fig:getupTimes}
\end{figure}

\subsection{Robust Recoveries}
During the pre-training stage, the low-level policy is trained to recover from random initial states, which leads the policy to develop robust recovery strategies that enable the character to consistent get back up after falling. These recovery strategies can then be conveniently transferred to new tasks by simply reusing the low-level policy. This allows the character to agilely execute recovery behaviors in order to get up after a fall, and then seamless transition back to performing a given task. These behaviors emerge automatically from the low-level policy, without requiring explicit training to recover from perturbations for each new task. Examples of the learned recovery strategies are available in Figure~\ref{fig:recoverySnapshots}. To evaluate the effectiveness of these recovery strategies, we apply random perturbation forces to the character's root and record the time required for the character to recover after a fall. Figure~\ref{fig:getupTimes} shows the amount of time needed for recoveries across 500 trials. A force between [500, 1000]N in a random direction is applied for 0.5s to the character's root. A fall is detected whenever a body part makes contact with the ground, excluding the feet. A successful recovery is detected once the character's root recovers back to a height above 0.5m and its head is above 1m. The policy is able to successfully recover from all falls, with a mean recovery time of 0.31s and a maximum recovery time of 4.1s. Even though the dataset does not contain any motion clips that depict get up behaviors, the low-level policy is still able to discover plausible recovery strategies, such as using its hands to break a fall and get back up more quickly. However, the character can still exhibit some unnatural recovery behaviors, such as using overly energetic spins and flips. Including motion data for more life-like recovery strategies will likely help to further improve the realism of these motions.

\section{Discussion and Future Work}
In this work, we presented adversarial skill embeddings, a scalable data-driven framework for learning reusable motor skills for physics-based character animation. Our framework enables characters to learn rich and versatile skill embeddings by leveraging large unstructured motion datasets, without requiring any task-specific annotation or segmentation of the motion clips. Once trained, the skill embeddings can be reused to synthesize naturalistic behaviors for a diverse array of downstream tasks. Users can specify tasks through simple high-level reward functions, and the pre-trained low-level policy then allows the character automatically utilize and compose life-like behaviors in furtherance of the task objectives.

Our system demonstrates that large scale adversarial imitation learning can be an effective paradigm for developing general-purpose motor skill models for a wide range of sophisticated behaviors. However, like many GAN-based techniques, our training procedure for the low-level policy is still prone to mode-collapse. While the skill discovery objective can greatly improve the diversity of behaviors learned by the model, the policy still cannot fully capture the rich variety of behaviors depicted in a large motion dataset. Exploring techniques to mitigate mode-collapse could lead to more expressive and versatile skill embeddings that enable characters to tackle more complex tasks, as well as to synthesize more graceful and agile behvaiors. The ASE objective detailed in Section~\ref{sec:ASE} provides a general pre-training objective, where different approximations can be used for each component. We proposed using a GAN-based approximation for the distribution matching objective, and a particular variational approximation for the mutual information objective. Alternative approximations can be used, such as flow models \cite{RealNVPDinh2017}, diffusion models \cite{dickstein15}, and contrastive predictive coding \cite{oord2018representation}, which can present trade-offs that provide a rich design space for future exploration. While our model is able to perform a large array of skills, the downstream tasks that we have explored in our experiments are still relatively simple. Applying ASE to more challenging tasks in complex environments that require composition of a wider array of skills will help to better understand the capabilities of these models. While our framework is able to leverage massively parallel simulators to train skill embeddings with years of simulated data, this process is highly sample intensive compared to the efficiency with which humans can explore and acquire new skills. We would be interested in exploring techniques that can better replicate the skill acquisitions of humans, which may improve the efficiency of the training process and further enhance the capabilities of the acquired skills. Finally, the low-level policy can still occasionally produce unnatural motions, such as jittering, sudden jerks, and overly energetic recovery behaviors. We are interested in exploring methods to mitigate these artifacts and further improve the realism of the generated motions, such as incorporating motion data for natural recovery strategies and integrating energy efficiency objectives into the pre-training stage \cite{2021-TOG-AMP,SymYu2018}. Despite these limitations, we hope this work will help pave the way towards developing large-scale and widely reusable data-driven control models for physics-based character animation.

\begin{acks}
We would like to thank \href{https://actorcore.reallusion.com/}{Reallusion\footnote{\href{https://actorcore.reallusion.com/}{actorcore.reallusion.com}}} for providing reference motion data for this project, the anonymous reviewers for their feedback, Charles Zhou, Dane Johnston, David Dixon, Simon Yuen, Viktor Makoviychuk, and Gavriel State for their support for this work.
\end{acks}
% acknowledgements: Reallusion, 

\bibliographystyle{ACM-Reference-Format}
\bibliography{ase}

\pagebreak

\appendix

\section{Hyperparameters}
\label{app:hyperparams}

Hyperparameter settings used during pre-training of the low-level policy are available in Table~\ref{tab:suppASELLPParams}, and the hyperparameters for task-training of the high-level policy are available in Table~\ref{tab:suppASEHLPParams}. During pre-training, the trajectories are generated by conditioning the low-level policy $\pi$ on a random sequence of latents $\rmZ = \{\rvz_0, \rvz_1, ..., \rvz_{T-1} \}$ sampled according to $p(\rvz)$. The sequence of latents is constructed such that a latent $\rvz$ is repeated for multiple timesteps, before a new latent is sampled from $p(\rvz)$ and repeated for multiple subsequent timesteps. Each latent is kept fixed for between 1 and 150 timesteps before being changed. During task-training, the high-level policy is queried at 6Hz, while the low-level policy operates at 30Hz. The latents specified by the high-level policy is therefore repeated for 5 steps for the low-level policy. This can help improve motion quality by lowering the rates at which the high-level policy can change the skill being executed by the low-level policy, reducing the prevalence of unnatural jittery behaviors.

\begin{table}[h!]
{ \centering  
\caption{ASE hyperparameters for training low-level policy.}
\label{tab:suppASELLPParams}
\begin{tabular}{|l|c|}
\hline
{\bf Parameter} & {\bf Value}  \\ \hline
    $\mathrm{dim}(\mathcal{Z})$ Latent Space Dimension &  $64$  \\ \hline
    $\Sigma_\pi$ Action Distribution Variance &  $0.0025$  \\ \hline
    $\beta$ Skill Discovery Objective Weight &  $0.5$  \\ \hline
    $w_\mathrm{gp}$ Gradient Penalty Weight &  $5$  \\ \hline
    $w_\mathrm{div}$ Diversity Objective Weight &  $0.01$  \\ \hline
    $\kappa$ Encoder Scaling Factor &  $1$  \\ \hline
    Samples Per Update Iteration &  $131072$  \\ \hline
    Policy/Value Function Minibatch Size &  $16384$  \\ \hline
    Discriminator/Encoder Minibatch Size &  $4096$  \\ \hline
    $\gamma$ Discount &  $0.99$  \\ \hline
    Adam Stepsize & $2 \times 10^{-5}$ \\ \hline
    GAE($\lambda$) &  $0.95$  \\ \hline
    TD($\lambda$) &  $0.95$  \\ \hline
    PPO Clip Threshold &  $0.2$  \\ \hline
    $T$ Episode Length &  $300$  \\ \hline
\end{tabular}
}
\end{table}

\begin{table}[h!]
{ \centering  
\caption{ASE hyperparameters for training high-level policy.}
\label{tab:suppASEHLPParams}
\begin{tabular}{|l|c|}
\hline
{\bf Parameter} & {\bf Value}  \\ \hline
    $w_G$ Task-Reward Weight &  $0.9$  \\ \hline
    $w_S$ Style-Reward Weight &  $0.1$  \\ \hline
    $\Sigma_\pi$ Action Distribution Variance &  $0.01$  \\ \hline
    Samples Per Update Iteration &  $131072$  \\ \hline
    Policy/Value Function Minibatch Size &  $16384$  \\ \hline
    $\gamma$ Discount &  $0.99$  \\ \hline
    Adam Stepsize & $2 \times 10^{-5}$ \\ \hline
    GAE($\lambda$) &  $0.95$  \\ \hline
    TD($\lambda$) &  $0.95$  \\ \hline
    PPO Clip Threshold &  $0.2$  \\ \hline
    $T$ Episode Length &  $300$  \\ \hline
\end{tabular}
}
\end{table}

\end{document}